\documentclass[a4,semcolor,psfig,english,12pt,epsf,portrait]{article}
\usepackage{amsmath,amsfonts,latexsym,amscd,amssymb,theorem}

\usepackage[T1]{fontenc}
\usepackage{epsfig}
\usepackage{amsmath}
\usepackage{psfrag}

\def\div{\hbox{div}}

\def\R{\hbox{\bf R}}
\def\Z{\hbox{\bf Z}}

\def\d{\displaystyle}

\def\div{\hbox{div}}

\def\N{\hbox{\bf N}}

\def\a{\alpha}

\def\D{{\cal D}}

\def\e{\varepsilon}

\def\<{\langle}
\def\>{\rangle}
\newcommand{\ba}{\begin{eqnarray}}
\newcommand{\ea}{\end{eqnarray}}

\newtheorem{theo}{\bf Theorem}[section]
\newtheorem{lem}[theo]{\bf \textit{Lemma}}
\newtheorem{pro}[theo]{\bf Proposition}
\newtheorem{cor}[theo]{\bf Corollary}
\newtheorem{defi}[theo]{\bf Definition}
\newtheorem{rem}[theo]{\bf Remark}

\renewcommand{\R}{{\mathbb R}}
\renewcommand{\Z}{{\mathbb Z}}
\renewcommand{\N}{{\mathbb N}}

\begin{document}

\title{\bf Global continuous solutions\\
to diagonalizable  hyperbolic systems\\
  with large and monotone data}

\author{
\normalsize\textsc{A. El
  Hajj$^1$$^2$, R. Monneau$^1$}}
\vspace{20pt}
\maketitle
\footnotetext[1]{\'Ecole Nationale des Ponts et
 Chauss\'ees, CERMICS,
6 et 8 avenue Blaise Pascal, Cit\'e Descartes
 Champs-sur-Marne, 77455 Marne-la-Vall\'ee Cedex 2, France}
\footnotetext[2]{ Université de Marne-la-Vallée
5, boulevard Descartes
Cité Descartes - Champs-sur-Marne
77454 Marne-la-Vallée cedex 2}


 \centerline{\small{\bf{Abstract}}}
 \noindent{\small{In this paper, we study  diagonalizable hyperbolic
     systems in one space dimension. Based on 
      a new  gradient entropy estimate, we prove the
      global existence of a continuous solution,   
      for large and nondecreasing initial data. Moreover, we show in particular cases
      some uniqueness results. We also remark that these results cover
       the case of systems which are hyperbolic but not strictly hyperbolic. Physically, this
      kind of diagonalizable hyperbolic systems appears naturally in the
      modelling of the dynamics of dislocation densities.
 }}

\hfill\break
 \noindent{\small{\bf{AMS Classification: }}} {\small{35L45, 35Q35,
     35Q72, 74H25.}}\hfill\break
  \noindent{\small{\bf{Key words: }}} {\small{Global existence, system of Burgers
      equations, system of nonlinear transport equations, nonlinear
      hyperbolic system, dynamics of dislocation densities.}}\hfill\break

\vspace{20pt}

\section{Introduction and main result}

\subsection{Setting of the problem}
In this paper we are interested in  continuous solutions to hyperbolic
systems in dimension one. Our work will focus on  
solution $\d{u(t,x)=(u^i(t,x))_{i=1,\dots,M}}$, where $M$ is an integer,
of hyperbolic systems which are
diagonal, i.e.
\begin{equation}\tag{P}\label{EM:burger}
\partial_t u^i + a^i(u)\partial_x u^i = 0 \quad \mbox{on} \quad  
(0,T)\times\R\quad \mbox{and for}\quad i=1,...,M,   
\end{equation}

\noindent with the initial data:
\begin{equation}\tag{ID}\label{EM:initialdata}
u^i(0,x)=u_0^i(x),\qquad \mbox{$x\in\mathbb{R}$, for $i=1,\dots,M$} .
\end{equation}

\noindent  For real numbers $\alpha^i \le \beta^i$, let us consider the box
\begin{equation}\label{EM:box}\displaystyle{U= \Pi_{i=1}^M [\alpha^i,\beta^i]}.
\end{equation}

\noindent We consider a given  function  
 $a=(a^i)_{i=1,...,M}: U\to \R^M$, which 
 satisfies the following regularity assumption:

$$(H1) \quad \left\{\begin{array}{l}
\mbox{the function $a\in C^{\infty}(U)$,}\\
\\
\mbox{there exists}\quad M_0 >0  \quad \mbox{such that
       for}\quad i=1,...,M,\\
|a^i(u)|\le M_0 \quad \mbox{for all}\quad
       u\in  U,\\
\\
\mbox{there exists}\quad M_1 >0  \quad \mbox{such that
       for}\quad i=1,...,M,\\
      
       |a^i(v)-a^i(u)| \le M_1 |v-u| \quad \mbox{for all}\quad
       v,u\in U.
\end{array}\right.$$

\noindent We assume, for all $u\in \R^M$, that the matrix

$$\mbox{$(a^i_{,j}(u))_{i,j=1,...,M}$, where
$\d{a^i_{,j}=\frac{\partial}{\partial u^j} a^i,}$}$$

\noindent is non-negative in the positive cone,
namely

$$(H2) \left|\begin{array}{l}
\mbox{for all}\quad u\in U,\quad
  \mbox{we have}\\
\\
  \displaystyle{\sum_{i,j=1,...,M}\xi_i\xi_j a^i_{,j}(u) \ge 0\quad \mbox{for
  every}\quad \xi=(\xi_1,...,\xi_M)\in [0,+\infty)^M.}
\end{array}\right.$$

\noindent In (\ref{EM:initialdata}), each component $u_0^i$ of
the initial data $u_0=(u_0^1,\cdots,u_0^M)$ is assumed satisfy the
following property:\\

$$(H3)\quad \left.\left\{\begin{array}{l}
\mbox{$u_0^i\in L^{\infty}(\mathbb{R})$,}\\
\\
\mbox{$u_0^i$ is nondecreasing,}\\
\\
\mbox{$\partial_x u_0^i \in L\log L(\R)$,}
      \end{array}\right.\right|\mbox{ for $i=1,\cdots,M$,}
$$

\noindent where  $L\log L(\R)$
is the following Zygmund space:

 $$L\log L(\R)=\left\{\mbox{$f\in L^1(\R)$ such that
  $\displaystyle{\int_{\R}|f|\ln\left(1+|f|\right)<+\infty}$}
 \right\}.$$
\noindent  This space is
equipped by the following norm:

$$\|f\|_{L\log
  L(\R)}=\inf\left\{\lambda>0:\displaystyle{\int_{\R}}\frac
  {|f|}{\lambda}\ln\left(1+\frac {|f|}{\lambda}\right)\le 1\right\},$$

\noindent  This norm is due to Luxemburg (see Adams \cite[(13), Page
234]{Adams}).\\

\noindent Our purpose is to show the
existence of a continuous solution, such that $u^i(t,\cdot)$ 
 satisfies  $(H3)$ for all time.

\subsection{Main result}

\noindent It is well-known that for the classical Burgers equation, the solution
stays continuous when the initial data is Lipschitz-continuous and
non-decreasing. We want somehow to generalize this result to the case of diagonal
hyperbolic  systems.

\begin{theo}\textbf{(Global existence of a nondecreasing solution)}\label{EM:th1}\\
Assume $(H1)$, $(H2)$ and $(H3)$. Then, for all
$T>0$, we have:\\

\noindent i) \underline{{\bf Existence of a weak solution:}}

\noindent There exists a function $u$ solution of (\ref{EM:burger})-(\ref{EM:initialdata}) (in the
distributional sense), where 
$$\mbox{$u\in [L^{\infty}((0,T)\times\R)]^M \cap [C([0,T);L\log
L(\R))]^M$ and  $\partial_x u \in [L^{\infty}((0,T);L\log
L(\R))]^M$},$$
\noindent  such that for a.e $t\in [0,T)$ the function $u(t,\cdot)$ is
nondecreasing in $x$  and  satisfies the following $L^{\infty}$ estimate:

\begin{equation}\label{EM:max_pri}
\|u^i(t,\cdot)\|_{L^{\infty}(\R)}\le \|u^i_0\|_{L^{\infty}(\R)}, \qquad \mbox{for
$i=1,\dots,M$},
\end{equation}
\noindent and the  gradient entropy estimate:
\begin{equation}\begin{array}{ll}\label{EM:entropy}
\displaystyle{
\int_{\R}\sum_{i=1,\dots,M}f\left(\partial_x
  u^i(t,x)\right)dx}+ 
\int_{0}^t\int_{\R} \displaystyle{\sum_{i,j=1,\dots,M}a^i_{,j}(u)\partial_x
u^i(s,x)\partial_x u^j(s,x)\;dx\;ds}\le C_1,
\end{array}\end{equation}

\noindent where 
\begin{equation}\label{EM:f}f(x)=\left\{\begin{array}{ll}
x\ln(x)+\frac{1}{e} & \quad \mbox{if}\quad x \ge 1/e,\\
0 & \quad \mbox{if}\quad 0\le x \le 1/e,\\
\end{array}\right.\end{equation}

\noindent and $C_1(T,M,M_1,\|u_0\|_{[L^{\infty}(\R)]^M},\|\partial_x u_0\|_{[L\log L
  (\R)]^M})$.\\ 

\noindent ii) \underline{{\bf  Continuity of the solution:}}

\noindent The solution $u$ constructed in (i) belongs to $C([0,T)\times
\R)$ and  there exists 
a modulus  of continuity $\omega(\delta,h)$,  such
that for all $(t,x)\in(0,T)\times \R$ and all $\delta,h\ge 0$, we have:

\begin{equation}\label{EM:contu}|u(t+\delta,x+h)-u(t,x)|\le C_2\;\omega(\delta,h)
\;\;\mbox{with}\;\; \displaystyle{\omega(\delta,h)=\frac{1}{\ln(\frac
    {1}{\delta}+1)}+\frac{1}{\ln(\frac {1}{h}+1)}}.
\end{equation}
\noindent where $C_2(T,M_1, M_0,\|u_0\|_{[L^{\infty}(\R)]^M},\|\partial_x u_0\|_{[L\log L
  (\R)]^M})$.
\end{theo}

\begin{rem}$\;$\\
Here,  we can easily 
extend the solution $u$ of (\ref{EM:burger})-(\ref{EM:initialdata}), given by
Theorem \ref{EM:th1}, on the time interval $[0,+\infty)$.
\end{rem}

\noindent Our method is based on the following simple remark: if the initial
data satisfies $(H3)$ then the
solution satisfies $(H3)$ for
all $t$. What seems  new is the gradient entropy inequality. 
The prove of Theorem \ref{EM:th1} is rather standard.
First we regularize the initial data and
the system with the addition of a viscosity term, then we show that
this regularized system admits a classical solution for short time. We prove the bounds
(\ref{EM:max_pri}) and the  fundamental  gradient entropy inequality
(\ref{EM:entropy}) which allow to get a solution for all time.
Finally,  these {\it a priori} estimates ensure enough compactness to
pass to the limit when the regularization varnishes  and to get the
existence of a solution.\\

\begin{rem}$\;$\\
To guarantee the $L\log L$ bound on  the gradient  of the solutions. 
We assumed in (H2) a sign on the left
hand side of gradient entropy inequality (\ref{EM:entropy}).
\end{rem}

\noindent In the case of $2\times2$ strictly hyperbolic systems,
which corresponds in (\ref{EM:burger}) to the case of
$a^1(u^1,u^2)<a^2(u^1,u^2)$. Lax
\cite{Lax} proved the existence of smooth solution of 
(\ref{EM:burger})-(\ref{EM:initialdata}). This result was also proven by 
Serre \cite[Vol II]{Serre12} in the case of $M\times M$ 
rich hyperbolic systems (see 
also Subsection \ref{EM:ref} for more related references). Their result is limited to the case of strictly
hyperbolic systems, here in  Theorem \ref{EM:th1}, we treated the case
of systems which are hyperbolic but not strictly hyperbolic. See the
following Remark for a quite detailed example.\\

\begin{rem}{\bf (Crossing eigenvalues)}\label{EM:cross}\\
 Condition (\ref{EM:str_hy}) on the
eigenvalues is required in our framework (Theorem \ref{EM:th1}). Here is a  
 simple example of a $2\times2$  hyperbolic but not strictly hyperbolic
 system. We consider solution $u=(u^1,u^2)$ of 

\begin{equation}\label{EM:exem}
\left.\left\{\begin{array}{ll}\partial_t u^1+cos(u^2)\partial_x u^1 =0,\\
\\
\partial_t u^2+u^1sin(u^2)\partial_x u^2
=0,\end{array}\right.\right|\;\;\mbox{on $(0,T)\times \R$.}
\end{equation}

\noindent Assume:\\ 

\noindent i)  $u^1(-\infty)=0$, $u^1(+\infty)=1$ and $\partial_x u^1\ge 0$,\\

\noindent ii) $u^2(-\infty)=-\frac{\pi}{2}$,
$u^2(+\infty)=\frac{\pi}{2}$ and $\partial_x u^2\ge 0$.\\

\noindent Here the eigenvalues
$\lambda_1(u^1,u^2)=cos(u^2)$ and $\lambda_2(u^1,u^2)=u^1sin(u^2)$
cross each other at the initial time (and indeed for any
time). Nevertheless for $a^1(u^1,u^2)=cos(u^2)$ and
$a^2(u^1,u^2)=u^1sin(u^2)$, we can compute

 $$(a^i_{,j}(u^1,u^2))_{i,j=1,2}=\left(\begin{array}{ccc}  0& -sin(u^2) \\
                              sin(u^2)& u^1cos(u^2)         
\end{array}\right),$$
\noindent which satisfies $(H2)$ (under assumptions (i) and (ii)). Therefor
Theorem \ref{EM:th1} gives the existence of a solution to (\ref{EM:exem}) with
(i) and (ii). 
\end{rem}

\noindent Based on the same type of gradient entropy inequality (\ref{EM:entropy}), it was
proved in  Cannone et al. \cite{EC} the
existence of a solution in the distributional sense for a 
two-dimensional system of two transport equations, where the  
velocity vector field is  non-local.\\

\noindent The uniqueness of the solution is strongly related to 
the existence of regular (Lipschitz) solutions (see
Theorem \ref{EM:unicite}). Let us remark that equation
(\ref{EM:burger})-(\ref{EM:initialdata}) does not create shocks because the
solution (given in Theorem \ref{EM:th1}) is continuous. In this situation, it seems
very natural to expect the uniqueness of the solution. Indeed the notion of
entropy solution (in particular designed to deal with the
discontinuities of weak solutions) does not seem so helpful in this
context. Nevertheless the uniqueness of the solution is an open problem
in general (even for such a simple system).\\

We ask the following {\it {\bf  Open question:}}

\noindent Is there  uniqueness of the solution given in Theorem
\ref{EM:th1} ?\\

\noindent Now we give the following existence and uniqueness result in
$[W^{1,\infty}([0,T)\times\R)]^M$, in a special  case to simplify the
presentation. More precisely we assume

$$\mbox{$(H1')\quad\d{a^i(u)=\sum_{j=1,\dots,M}A_{ij}u^j}$ for
  $i=1,\dots,M$ and for all
$u\in U$,}$$
 
$$\hspace{1cm}(H2')\quad
  \displaystyle{\sum_{i,j=1,...,M}A_{ij}\xi_i\xi_j \ge 0\quad \mbox{for
  every}\quad \xi=(\xi_1,...,\xi_M)\in [0,+\infty)^M.}$$

\begin{theo}{\bf (Existence and uniqueness of $W^{1,\infty}$  solution
    for a particular $A=(A_{ij})_{i,j=i=1,\dots,M}$)}\label{EM:unicite1}\\
Assume $(H1')$. For $T> 0$ and all nondecreasing initial data $u_0\in
[W^{1,\infty}(\R)]^M$, the system
(\ref{EM:burger})-(\ref{EM:initialdata}) admits a unique solution $u\in
\left[W^{1,\infty}([0,T)\times\R)\right]^M$, in the following cases:\\

\noindent i)  $M\ge 2$ and  $A_{ij}\ge 0$, for all
$j\ge i$.

\noindent ii) $M\ge 2$ and $A_{ij}\le 0$,  for all
$i\neq j$ and $(H2')$. And then for all $(t,x)\in [0,T)\times \R$ we have

\begin{equation}\label{EM:w_infty}
\sum_{i=1,\dots,M}\partial_x u^i(t,x)\;\;\le \;\;\sup_{y\in
  \R}\sum_{i=1,\dots,M}\partial_x u^i_0(y).
\end{equation}
\end{theo}

\begin{rem}{\bf (Case of $M=2$)}\\
In particular  for  $M=2$, if $(H1')$, $(H2')$ and
$(H3)$ satisfied then we have, by Theorem \ref{EM:unicite1} 
the existence and uniqueness of a solution in
$\left[W^{1,\infty}([0,T)\times\R)\right]^2$
 of (\ref{EM:burger})-(\ref{EM:initialdata}).
\end{rem}

\noindent In these particular cases  of the matrix $A$, we can prove
that $\partial_x u^i$ for $i=1,\dots,M$, are bounded on $[0,T)\times\R$. Thanks to this better
estimates on  $\partial_x u^i$, and then on the velocity  vector
field $Au$,  we prove here the uniqueness of the solution. \\

\noindent In the case of the matrix 
$A=\left(\begin{array}{ccc}  1& -1 \\
                              -1& 1          
\end{array}\right)$, 
it was proved in El Hajj,  Forcadel \cite{EF}, the existence and uniqueness of a
Lipschitz viscosity solution, and in A. El Hajj \cite{EL}, the existence
and uniqueness of a strong solution in $W^{1,2}_{loc}([0,T)\times
\R)$.\\

\subsection{Application to diagonalizable systems}\label{EM:dia}
Let us first consider a smooth function $u=(u^1,\dots,u^M)$, solution of
the following non-conservative hyperbolic system:
\begin{equation}\label{EM:lef}
\left\{\begin{array}{ll}\partial_t u(t,x)+F(u)\partial_x u(t,x)=0,
&u(t,x)\in U,\;x\in\R,\; t\in (0,T),\\
\\
u(x,0)=u_0(x) &x\in\R,\end{array}\right.\end{equation}

\noindent  where the
space of states $U$ is now an open subset of $\R^M$, and for each $u$, $F(u)$ is a
$M\times M$-matrix and the  map $F$ is of class $C^1(U)$.  We assume
that  $F(u)$ has $M$ real eigenvalues $\lambda_1(u),\dots,\lambda_M(u)$,
and we suppose that we can select bases of right and left eigenvectors
$r_i(u)$, $l_i(u)$ normalized so that

$$|r_i|\equiv 1 \;\;\;\mbox{and}\;\;\; l_i\cdot r_j=\delta_{ij} $$
\begin{rem}{\bf (Riemann invariant)}\\
Recall that locally a necessary and sufficient condition to write 
$$l_i(u)=\nabla_u \varphi_i (u),$$

\noindent is the Frobenius condition $l_i \wedge dl_i=0$. In that case
the function $\varphi_i(u)$ is solution of the following equation

$$(\varphi_i(u))_t + \lambda_i(u)(\varphi_i(u))_x =0.$$

\noindent We recall that then
$\varphi_i(u)$ is called a $i$-Riemann invariant (see Sevennec
\cite{Sevenn} and Serre \cite[Vol II]{Serre12})). If this is true for
any $i$, we say that the system (\ref{EM:lef})
is diagonalizable.
\end{rem}

\noindent Our theory is naturally applicable to this more general class
of systems.

\subsection{A brief review of some related literature}\label{EM:ref}
\noindent Now we  recall some well known results for system
(\ref{EM:lef}).

\noindent For a scalar conservation law, this corresponds in (\ref{EM:lef}) to the case
$M=1$ and $F(u)= h'(u)$ is the derivative of some flux function $h$, the
global existence and uniqueness of $BV$ solution established by Oleinik
\cite{Ole} in one space dimension. The famous paper of Kruzhkov
\cite{Kru} covers the more general class of $L^{\infty}$ solutions, in
several space dimension. For  another alternative approach  based on the
notion of entropy process solutions, see Eymard et
al. \cite{Eymard}, see also the kinetic formulation P. L. Lions et
al. \cite{LBT}. \\

\noindent We now recall some well-known results for 
a class of $2\times 2$ strictly hyperbolic systems n one space
dimension. Here i.e
$F(u)$ has two real, distinct eigenvalues 

$$\lambda_1(u)< \lambda_2(u).$$

\noindent Lax \cite{Lax} proved the existence and  uniqueness of
nondecreasing  and smooth solutions
of the $2\times 2$  strictly hyperbolic systems. Also in case of
$2\times 2$ strictly hyperbolic systems DiPerna \cite{DiPerna2,
  DiPerna3}  showed the global existence of a $L^{\infty}$ solution. The
proof of DiPerna relies on a
compensated compactness argument, based on the representation of the
weak limit in terms of Young measures, which must reduce to a Dirac mass
due to the presence of a large family of entropies. This results is
based on the idea of Tartar \cite{Tartar}.\\

\noindent For general  $M\times M$  strictly
hyperbolic systems; i. e. where $F(u)$ has $M$ real,  distinct eigenvalues

\begin{equation}\label{EM:str_hy}\lambda_1(u)< \dots<\lambda_M(u),\end{equation}

\noindent  Bianchini and Bressan proved in  \cite{Bressan} a  striking global
existence and uniqueness result of $BV$ solutions to system (\ref{EM:lef}),
assuming that the initial data has small total variation. Their existence
result is a generalization of Glimm result \cite{Glimm}, proved in
the conservation case; i.e. $F(u)=Dh(u)$ is the Jacobin of some flux
function $h$ and generalized by LeFloch and Liu \cite{LEF88,LEF93} in the
non-conservative case.\\

\noindent We can also mention that, our system (\ref{EM:burger}) is related
to other similar  models, such as scalar transport equations based on
vector fields with low regularity. Such equations were for
instance studied by Diperna and Lions in \cite{Dep}. They have proved
the existence (and uniqueness) of a solution (in the renormalized
sense), for given vector fields in
$L^1((0,+\infty);W^{1,1}_{loc}(\R^N))$ whose divergence is in
$L^1((0,+\infty); L^{\infty}(\R^N))$. This study was generalized
by Ambrosio \cite{Amb2004}, who considered vector fields in
$L^1((0,+\infty);BV_{loc}(\R^N))$ with bounded divergence. In the
present paper, we work in dimension $N=1$ and prove the existence
(and some uniqueness results) of solutions of the system
(\ref{EM:burger})-(\ref{EM:initialdata}) with a velocity  vector
field $a^i(u)$, $i=1,\dots,M$. Here, in Theorem \ref{EM:th1},  
 the divergence of our  vector field is only in $L^{\infty}((0,+\infty),
L\log L(\R))$. In this case we proved the existence result   
thanks to the gradient entropy estimate (\ref{EM:entropy}), which gives a better estimate
on the solution. However, in  Theorem \ref{EM:unicite1}, the divergence of
our vector field is bounded, which allows us to get a
uniqueness result for the non-linear system (\ref{EM:burger}).\\

\noindent We also refer to Ishii, Koike \cite{IK91} and  Ishii \cite{Ish92}, who showed existence
and uniqueness of viscosity solutions for Hamilton-Jacobi systems
of the form:
\begin{equation}\label{EM:ichi}\left\{\begin{array}{ll}
\partial_tu^i+ H_i(u,Du^i)=0&\mbox{with } \;u=(u^i)_i\in \R^M,\;\mbox{for } \;x\in\R^N,\; t\in (0,T),\\
\\
u^i(x,0)=u^i_0(x)&x\in\R,\end{array}\right.\end{equation} where
the Hamiltonian $H_i$ is quasi-monotone in $u$ (see
Ishii, Koike \cite[Th.4.7]{IK91}). This does not cover our study since
our Hamiltonian is not necessarily quasi-monotone.\\

\noindent For hyperbolic and symmetric systems, G$\dot{a}$rding 
has proved in \cite{Garding} a local existence and uniqueness
result in $C([0,T); H^s(\R^N))\cap C^1([0,T); H^{s-1}(\R^N))$,
with  $s> \frac N2+1$ (see also Serre \cite[Vol I, Th 3.6.1]{Serre12}), 
this result being only local in time, even in dimension $N=1$.
\subsection{Miscellaneous extensions to explore  in a futur work}

\noindent {\bf 1.} In  Theorem \ref{EM:th1} we have considered
the study of a particular system only  to simplify the presentation. This
result could be generalized to the following system

\begin{equation}\tag{P'}\label{EM:bur1}
\partial_t u^i + a^i(u,x,t)\partial_x u^i = h^i(u,x,t) \quad \mbox{on} \quad  
(0,T)\times\R\quad \mbox{and for}\quad i=1,...,M,   
\end{equation}

\noindent with suitable conditions on $a^i$ and $h^i$.\\

\noindent {\bf 2.} If we consider the case where the system
(\ref{EM:burger}) is strictly hyperbolic. Based in the result of  Bianchini,
Bressan \cite{Bressan},  we could also prove  the
uniqueness of the solution, whose existence is given by Theorem \ref{EM:th1}.

\noindent {\bf 3.} We could also extend  Theorem \ref{EM:unicite1} to
system (\ref{EM:bur1}), where we replace (i) and (ii) by the following condition\\

\noindent i') For $M\ge 2$, $a^i_j(u,x,t) \ge0$ for $j\ge i$ and for all $(u,x,t)\in
U\times \R\times [0,T)$.

 \noindent ii') For $M\ge 2$,
$$a^i_{,j}(u,x,t)\le 0 \quad \mbox{for all}\quad (u,x,t)\in U\times
\R\times [0,+\infty), \quad \mbox{for all} \quad i\not= j,$$
 \noindent and we assume that for any $v_i\in\R^M,x_i\in \R$, the matrix
$$b_{ij}(t)=a^i_{,j}(v_i,x_i,t)$$
satisfies for all $t\ge 0$
$$(H2'')\quad \d{\sum_{i,j=1,...,M}b_{ij}(t)}\xi_i\xi_j\ge 0 \quad \mbox{for all}\quad
\xi=(\xi_1,...,\xi_M)\in  [0,+\infty)^M.$$

\noindent {\bf 4.} We could also prove the uniqueness result in case of
$W^{1,\infty}$ solution among weak solution. (and in particular any weak
solution is a viscosity solution in the sense of  Crandall-Lions
\cite{Lio81, CL82}).\\

\noindent {\bf 5.} We could propose a numerical scheme and try
 to prove its convergence.\\

\noindent {\bf 6.} Applications to other equations: Euler, $p$-systems.
\subsection{Organization of the  paper}
This paper is organized as follows: in the Section 2, we approximate the
system (\ref{EM:burger})  and the initial conditions. Then we prove a local
in time existence for this  approximated system. In Section 3, we prove the global in time
existence for the approximated system. In the  Section 4, we
 prove that the obtained solutions are regular
and non-decreasing in $x$ for all $t\in(0,T)$. In Section 5, we 
prove the gradient entropy inequality and some other 
$\e$-uniform {\it a priori} estimates. In Section 6, we prove the 
main result (Theorem \ref{EM:th1}) passing to the limit as $\e$ goes to $0$ and using some
compactness properties inherited from our entropy  gradient inequality and the
{\it a priori} estimates. In Section 7 we prove some
uniqueness results in particular cases (Theorem \ref{EM:unicite1}). An application to the
dynamics of dislocation densities given in Section 8. Finally, in the
Appendix, we recall the proof of uniqueness of Lipschitz solution to
system (\ref{EM:burger}).

\section{Local existence of an approximated  system}

The system (\ref{EM:burger}) can be written as:
\begin{equation}\label{EM:burgers}
\partial_t u+ a(u) \diamond \partial_x u=0,
\end{equation}
where $u:=(u^i)_{1, \dots, M}$, $a(u)=(a^i(u))_{1, \dots, M}$
 and $U\diamond V$ is the
``component by component product'' of the two vectors
$U,V\in\mathbb{R}^M$. This is the vector in $\mathbb{R}^M$ whose
coordinates are given by $(U\diamond V)_i:=U_i V_i$:
$$\left[\begin{array}{l}U_1\\ U_2\\ \vdots\\
U_M\end{array}\right]\diamond \left[\begin{array}{l}V_1\\ V_2\\
\vdots\\ V_M\end{array}\right]=
\left[\begin{array}{l}U_1V_1\\ U_2V_2\\ \vdots\\
U_MV_M\end{array}\right].$$

\noindent Now, we consider  the system (\ref{EM:burgers}), modified by the
term $\e\partial_{xx}u$, where
$\displaystyle{\partial_{xx}=\frac{\partial^2}{\partial x^2}}$, 
 and for smoothed data. This modification
brings us to study, for all $0<\e\le 1$, the following system:

\begin{equation}\tag{$P_\e$}\label{EM:burgersapp}
\partial_t u^\e-\e\partial_{xx}u^\e=-a(u^\e) \diamond \partial_x u^\e,
\end{equation}

\noindent with the smooth initial data:

\begin{equation}\tag{$ID_\e$}\label{EM:initialapp}
u^{\e}(x,0)=u^{\e}_0(x),\;\;\; \mbox{ with $u^{\e}_0(x):=u_0\ast\eta_{\e}(x)$,}
\end{equation}

\noindent where $\eta_{\e}$  is a mollifier verify, $\eta_{\e}(\cdot)=\frac
{1}{\e}\eta(\frac{\cdot}{\e})$, such that $\eta\in
C^{\infty}_c(\R)$ is a non-negative function and  $\int_{\R}\eta=1$.

\begin{rem}{$\;$}\\ 
By classical properties of the  mollifier  $(\eta_{\e})_{\e}$ and the fact that
  $u_0^\e\in [L^{\infty}(\R)]^M$, then $ u_0^{\e}\in
  [C^{\infty}(\R)]^M\cap [W^{m,\infty}(\R)]^M$ for all  $m\in \N$.
\end{rem}

\noindent The global existence of smooth solution of the system
(\ref{EM:burgersapp}) is standard. Here, we prove this results only to
ensure the reader.\\

\noindent The following theorem is a local existence result (in the
"Mild" sense) of the regularized system  (\ref{EM:burgersapp})-(\ref{EM:initialapp}). This result
is achieved in a super-critical space. Here particularly we  chose
the space of functions $\left[C([0,T); X(\R))\right]^M$, where 
\begin{equation}\label{EM:X_R}X(\R)=\{\mbox{$u\in L^{\infty}(\R)$ such that $\partial_x u \in
  L^8(\R)$}\}.\end{equation}
\noindent  This space is a Banach
space supplemented with the  following norm

$$\|u\|_{X(\R)}=\|u\|_{L^{\infty}(\R)}+\|\partial_x u\|_{L^8(\R)}.$$

\noindent Here the espace $L^p(\R)$ with $p=8$ will simplify later in
Lemma \ref{EM:reg} the Bootstrap argument to get smooth solution.\\

\noindent In this Section, we will prove the following

\begin{theo}\label{EM:theo:exip}{\bf (Local existence result)}\\
For all initial data $u_0^\e\in [X(\R)]^M$ there exists
$$T^\star=T^\star(M_0,\e)>0,$$
such that the system  (\ref{EM:burgersapp})-(\ref{EM:initialapp}) admits
solutions $u^{\e}\in \left[C([0,T^{\star}); X(\R))\right]^M$.
\end{theo}

\noindent In order to do the proof of Theorem \ref{EM:theo:exip} in
Subsection \ref{EM:preu} we need to
recall in the following Subsection some known results.
\subsection{Useful results}
\begin{lem}\label{EM:eq:int}{\bf (Mild solution)}\\
Let $T>0$, and  $u^{\e}\in
\left[C([0,T); X(\R))\right]^M$ be a  solution of the
following integral problem with
$u^{\varepsilon}(t)=u^{\varepsilon}(t,\cdot)$:

\begin{equation}\tag{$IN_\e$}\label{EM:eq:i:6}
\displaystyle{u^{\varepsilon}(t)}=S_{\e}(t)u^{\e}_0
-\displaystyle{\int_0^t}S_{\e}(t-s)\left(a(u^\e(s))\diamond \partial_x
  u^\e(s)\right)ds,
\end{equation}

\noindent where $S_{\e}(t)=S_{1}(\e t)$ such that $S_{1}(t)=e^{t\Delta }$ is the
heat semi-group. Then $u^{\e}$ is a solution of the system 
(\ref{EM:burgersapp})-(\ref{EM:initialapp}) in the sense of distributions.
\end{lem}
For the proof of this lemma, we refer to Pazy \cite[Th 5.2. Page 146]{Pazy}.

\begin{lem}\label{EM:l2}\textbf{(Picard Fixed Point Theorem, see \cite{Vasile})}\\
Let $E$ be a Banach space, let  $B: E\times E \longrightarrow E$ be a
continuous  map such that:

$$\|B(x,y)\|_{E}\leq\eta\|y\|_{E}\;\;\;\mbox{for
  all}\;\;\;x,y\in E, $$
where $\eta$ is a positive given constant. Then, for
every $x_0\in E$, if 
$$0<\eta<1,$$
the equation $x=x_0+B(x,x)$ admits a solution in $E$. 
\end{lem}
\noindent In order to show the local existence of a solution for
(\ref{EM:eq:i:6}), we will apply  Lemma \ref{EM:l2} in  the space
$E=\left[L^{\infty}((0,T); X(\R))\right]^M$.

\begin{lem}\label{EM:Ctemps}{\bf(Time continuity)}\\
Let $T>0$. If $u^\e\in [L^{\infty}((0,T);W^{1, p}(\R))]^M$, $1\le p\le
+\infty$, are
solutions of integral problem (\ref{EM:eq:i:6}), 
then  $u^{\e}\in\left[C([0,T); W^{1,p}(\R))\right]^M$.
\end{lem}
For the proof of Lemma \ref{EM:eq:int}, see A. Pazy \cite[7.3, Page 212]{Pazy}.

\begin{lem}\label{EM:estsemi}{\bf(Semi-group estimates)}\\
Let $1\le p\le q \le +\infty$. Then for all $f\in L^{p}(\R)$ 
and for all $t>0$, we have the following estimates:\\

\noindent i) $\| S_{\e}(t)f\|_{L^{q}(\R)}\leq C t^{-\frac 12 (\frac 1p - \frac 1q)}\|f\|_{L^{p}(\R)}$,\\
 
\noindent ii) $\left\|\partial_x S_{\e}(t)f\right\|_{L^{p}(\R)}\leq C
t^{-\frac 12}\|f\|_{L^{p}(\R)},$\\

\noindent where $C=C(\e)$ is a positive constant depending on  $\e$.
\end{lem}
For the proof of this Lemma, see  Pazy \cite[Lemma 1.1.8, Th 6.4.5]{Pazy}.
\subsection{ Proof of  Theorem \ref{EM:theo:exip}}\label{EM:preu}

Our goal is to  show local existence  of a solution of
(\ref{EM:burgersapp}) using the Picard  fixed point Theorem. 	
To be done according  Lemma \ref{EM:eq:int} it is enough to prove the
local existence for the following equation:  
\begin{equation}\label{EM:int}\begin{array}{ll}
u^{\varepsilon}(t)
&=S_{\e}(t)u^{\e}_0
-\displaystyle{\int_0^t}S_{\e}(t-s)\left(a(u^\e(s))\diamond \partial_x
  u^\e(s)\right)ds,
\\
\\
&=S_{\e}(t)u^{\e}_0+B(u^\e,u^\e)(t),
\end{array}\end{equation}

\noindent with $B(u,v)(t)=-\displaystyle{\int_0^t}S_{\e}(t-s)\left(a(u) (s)\diamond \partial_x
  v(s)\right)ds$.\\

\noindent If we estimate $B(u,v)$, we will obtain, for all $u,v\in
[L^{\infty}((0,T); X(\R))]^M$, where $X(\R)$ defined in (\ref{EM:X_R}), the following:

\begin{equation}\label{EM:est_bi}\begin{array}{ll}\|B(u,v)(t)\|_{[X(\R)]^M}
&=\left\|\displaystyle{\int_0^t}
S_{\e}(t-s)\left(a(u (s))\diamond \partial_x
  v(s)\right)ds,\right\|_{[L^{\infty}(\R)]^M},\\
\\
&+\left\|\displaystyle{\int_0^t}
\partial_x S_{\e}(t-s)\left(a(u (s))\diamond \partial_x
  v(s)\right)ds,\right\|_{[L^{8}(\R)]^M},\end{array}\end{equation}

\noindent where for a function $f=(f^1,\dots,f^M)\in [X(\R)]^M $, we
note here 

$$\displaystyle{\|f\|_{[X(\R)]^M}=\sup_{i=1,\dots,M}\|f^i\|_{L^{\infty}(\R)}+\sup_{i=1,\dots,M}\|\partial_x
f^i\|_{L^{8}(\R)}}.$$

\noindent Using Lemma   \ref{EM:estsemi} (i) with $p=8, q=\infty$ for the first term  and Lemma
 \ref{EM:estsemi} (ii)  with $p=8$ for the second term,
 we obtain that :
$$\begin{array}{ll}\|B(u,v)(t)\|_{[X(\R)]^M}
&\hspace{-0.3cm}\leq
C\displaystyle{\int_0^t\frac{1}{(t-s)^{\frac{7}{16}}}}
\left\|a(u(s))\partial_x v(s)\right\|_{[L^{2}(\R)]^M}ds,\\
\\
&+C\displaystyle{\int_0^t\frac{1}{(t-s)^{\frac{1}{2}}}}
\left\|a(u(s))\partial_x v(s)\right\|_{[L^{8}(\R)]^M}ds.
\end{array}$$

\noindent We use the Hölder inequality, and get, for all $0<T\le 1$:

\begin{equation}\begin{array}{ll}\label{EM:bil}\|B(u,v)(t)\|_{[X(\R)]^M}
&\leq
C T^{\frac 12} \left\|\partial_x
  v\right\|_{[L^{\infty}((0,T); L^{8}(\R))]^M},\\
\\
&\leq
C T^{\frac 12}  \left\|
  v\right\|_{[L^{\infty}((0,T); X(\R))]^M},
\end{array}\end{equation}

\noindent where $C(M_0,\e)$. Moreover, we know by classical properties of heat semi-group
(see A. Pazy \cite{Pazy}):

\begin{equation}\label{EM:eq:do}
\|S_{\e}(t)u_{0}^{\e}\|_{[L^{\infty}((0,T); X(\R))]^
M} \le \|u_0^{\e}\|_{[X(\R)]^M} .
\end{equation}

\noindent Now, taking

\begin{equation}\label{EM:de_te}\displaystyle{(T^{\star})^{\frac
      12}=\min\left(\frac {1}{2C},1\right)},\end{equation}

\noindent we can easily verify that 
$$C(T^{\star})^{\frac 12}< 1.$$
\noindent By applying the Picard  Fixed Point Theorem (Lemma \ref{EM:l2}) with
$E=[L^{\infty}((0,T^\star); X(\R))]^M$, this proves the existence of a solution
$u^\e\in[L^{\infty}((0,T^\star); X(\R))]^M$ for (\ref{EM:int}).\\  

\noindent Then, according to Lemma \ref{EM:Ctemps}, we deduce that the
solution is indeed in  $[C([0,T^\star); X(\R))]^M$.\\

\noindent This proves, by Lemma  \ref{EM:eq:int}, the existence of a
solution  in $[C([0,T^\star); X(\R))]^M$, which satisfies the system
 (\ref{EM:burgersapp})-(\ref{EM:initialapp})  in the sense of distributions.
 $\hfill\Box$

\section{Global existence of the solutions of the approximated  system}

In this Section, we will prove the global existence of solution for the system
(\ref{EM:burgersapp})-(\ref{EM:initialapp}). Before going into the proof, we
need the following lemma.

\begin{lem}\label{EM:Ldeu}{\bf($L^{\infty}$ bound)}\\
 Let $T>0$. If $u^{\e}\in [C([0,T); X(\R))]^M$
is a solution of system (\ref{EM:burgersapp})-(\ref{EM:initialapp}) with
initial data $u_0^\e\in X(\R)$, then
  $$\left\|u^{\varepsilon}\right\|_{\left[L^{\infty}([0,T)\times\mathbb R)\right]^M}\leq
\|u_0^\e\|_{\left[ L^{\infty}(\R)\right]^M}$$
\end{lem}
\noindent The proof of this Lemma is a direct application of the
Maximum Principle Theorem for parabolic equations  (see
Gilbarg-Trudinger \cite[Th.3.1]{CI-TH}).

\begin{rem}\label{EM:M_0}$\;$\\
Thanks to the previous Lemma, we notice that we can take
the box $U$ defined in (\ref{EM:box}) as the following
$$\displaystyle{U= \Pi_{i=1}^M
  [-\|u_0^{\e,i}\|_{L^{\infty}(\R)},\|u_0^{\e,i}\|_{L^{\infty}(\R)}]}.$$

\noindent For fixed $\e$, this definition guarantee that $M_0$ do not change in
the course of time.
\end{rem}

\noindent The result of this Section is the following.

\begin{theo}\label{EM:theo:exip1}{\bf (Global existence)}\\
Let $T>0$ and $0<\e\le 1$. For initial data
$u_0^\e\in \left[X(\R)\right]^M$ satisfying $(H1)$ and $(H2)$. Then  the system
(\ref{EM:burgersapp})-(\ref{EM:initialapp}), admits a solution
$u^{\e}\in [C([0,T); X(\R))]^M$, with
$u^{\e}(t,\cdot)$  satisfying  $(H1)$ and $(H2)$
for all  $t\in (0,T)$. Moreover, for all  $t\in (0,T)$, we have the
following inequalities: 
\begin{equation}\label{EM:max_pri_e}
\|u^{\e, i}(t,\cdot)\|_{L^{\infty}(\R)}\le \|u^{\e,i}_0\|_{L^{\infty}(\R)}, \qquad \mbox{for
$i=1,\dots,M$},
\end{equation}
\end{theo}

\noindent {\bf Proof of Theorem \ref{EM:theo:exip1}:}\\
\noindent  We are going to prove that local in time  solutions obtained by
Theorem \ref{EM:theo:exip} can be extended to global
solutions  for the same system.\\

\noindent We argue by contradiction: assume that there exists a maximum
time $T_{max}$ such that, we have the existence of solutions of the
system  (\ref{EM:burgersapp})-(\ref{EM:initialapp}) in the function space $
[C([0,T_{max}); X(\R))]^M$. 
 
\noindent For every small enough $\delta>0$, we consider the system  (\ref{EM:burgersapp})
with the initial condition
$$u^{\e,\delta}_{0}(x)=u^{\e}(T_{max}-\delta,x).$$ 

\noindent From   Theorem \ref{EM:theo:exip} to deduce that there
exists a time $T^{\star}(M_0,\e)$, independent of $\delta$ (see Remark
\ref{EM:M_0}), such that the system (\ref{EM:burgersapp}) with initial data
$u^{\e,\delta}_{0}$ has a solution $u^{\e,\delta}$ on the time interval
$[0,T^{\star})$. Then for 

$$T_0=(T_{max}-\delta)+T^{\star},$$

\noindent we extend $u^\e$ on the time interval  $[0,T_0)$ as follows,

\begin{equation}
\tilde u^\e(t,x)=\left\{
\begin{aligned}
&u^\e(t,x),\quad\mbox{for}\quad t\in [0,T_{max}-\delta],\\
&u^{\e,\delta}(t,x),\quad\mbox{for}\quad t\in [T_{max}-\delta,T_0)
\end{aligned}
\right.\nonumber
\end{equation}

\noindent and we can check that $\tilde{u}^\e$ is a solution of
(\ref{EM:burgersapp})-(\ref{EM:initialapp}) on the time interval
$[0,T_0)$. But from Lemma (\ref{EM:Ldeu}) we know that the time $T^\star$ is
independent of $\delta$ (see Remark \ref{EM:M_0}), which implies that
$T_0>T_{max}$ and so a contradiction.

\noindent The inequalities (\ref{EM:max_pri_e}) 
  is a consequence of Lemma \ref{EM:Ldeu}. $\hfill\Box$


\section{Properties of the solutions of the approximated  system}

In this section, we are going to prove that the solution of 
(\ref{EM:burgersapp})-(\ref{EM:initialapp})  obtained by Theorem
\ref{EM:theo:exip} is smooth and monotone.

\begin{lem}\label{EM:reg}{\bf(Smoothness of the solution)}\\
 Let $T>0$. For all initial data  $u_0^\e\in [X(\R)]^M$, where
 $\partial_x u_0^\e\in [W^{m,p}(\R)]^M$ for all $m\in\N$, $1\le p\le +\infty$.\\
  
\noindent If $u^{\e}$ is a solution of the system 
(\ref{EM:burgersapp})-(\ref{EM:initialapp}), such that $u^{\e}\in\left[C([0,T); X(\R))\right]^M$ and
 $\partial_x u^{\e}\in [L^{\infty}((0,T);L^1(\R))]^M$, then $u^{\e}\in
 \left[C^{\infty}([0,T)\times \R)\right]^M$ and satisfies,

\begin{equation}\label{EM:requla}\displaystyle{ u^\e
    \in \left[W^{m,p}((0,T)\times\R)\right]^M
,\;\;\mbox{for all  $1< p \le +\infty$ and $m\in
  \N\setminus\{0\}$},}\end{equation}

\end{lem}
\noindent{\bf Proof of Lemma \ref{EM:reg}}\\

\noindent \underline{{\bf Step 1 (Initialization of the Bootstrap)}:}\\

\noindent For the sake of simplicity, we will set

$$F[u^\e]=-a(u^\e)\diamond \partial_x u^\e.$$ 

\noindent From the fact that $u^{\e}\in\left[C([0,T); X(\R))\right]^M$ and
$\partial_x u^{\e}\in [L^{\infty}((0,T);L^1(\R))]^M$, we deduce that
$\partial_x u^{\e},\;\;F[u^\e]\in \left[L^{1}((0,T)\times\R)\right]^M \cap
\left[L^{8}((0,T)\times\R)\right]^M$, which proves by interpolation that  

\begin{equation}\label{EM:etap0}\partial_x u^{\e},\;\;F[u^\e]\in
  \left[L^{p}((0,T)\times\R)\right]^M\;\;\;
\mbox{for all $1\le p \le 8$}.\end{equation}

\noindent Because $u^\e$ is a solution of (\ref{EM:burgersapp}), we see
that

\begin{equation}\label{EM:h_1}\partial_t u^\e-\e\partial_{xx}u^\e=F[u^\e],
\end{equation}

\begin{equation}\label{EM:h_2}\partial_{tx}
  u^\e-\e\partial_{xxx}u^\e=\partial_x F[u^\e].
\end{equation}

\noindent  Applaying the classical regularity theory of heat
equations on (\ref{EM:h_1}), we deduce that:  

\begin{equation}\label{EM:etap1}\partial_t u^\e \;\;\;\mbox{and}\;\;\;
  \partial_{xx} u^\e \in
  \left[L^{p}((0,T)\times\R)\right]^M,\;\;\mbox{for all $1 <p \le 8$}.
\end{equation}

\noindent For more details, see  Ladyzenskaja \cite[Theorem
9.1]{LSU}. But we know that

\begin{equation}\label{EM:D1}\partial_x F[u^\e]=-a(u^\e)\diamond \partial_{xx}
u^\e-Da(u^\e)\partial_x u^\e \diamond \partial_x u^\e\end{equation}

\noindent We notice that thanks to this better regularity on $u^\e$
((\ref{EM:etap0}) and (\ref{EM:etap1}),  and  by the Hölder inequality we can
easily prove that 

$$\partial_x F[u^\e]\in
\left[L^{p}((0,T)\times\R)\right]^M\;\;\;\mbox{for all $1<p\le 4$}.$$

\noindent Now, we apply again the classical regularity theory of heat
equations on (\ref{EM:h_2}), to deduce that:

\begin{equation}\label{EM:etap2}\partial_{tx} u^\e \;\;\;\mbox{and}\;\;\; 
\partial_{xxx} u^\e \in \left[
 L^{p}((0,T)\times\R)\right]^M,\;\;\mbox{for all  $1 <p \le
 4$}.\end{equation}

\noindent We know that

\begin{equation}\label{EM:D2}\partial_t F[u^\e]=-a(u^\e)\diamond \partial_{tx}
u^\e-Da(u^\e)\partial_t u^\e \diamond \partial_x u^\e\end{equation}

\noindent Thanks this previous regularity on $u^\e$, we obtain by the
Hölder inequality that 

$$\partial_t F[u^\e]\in
\left[L^{p}((0,T)\times\R)\right]^M\;\;\;\mbox{for all $1<p\le 4$}.$$

\noindent Which gives  that

$$\partial_{x}u^\e,\;\;F[u^\e]\in
\left[W^{1,p}((0,T)\times\R)\right]^M\;\;\;\mbox{for all $1<p\le 4$},$$

\noindent and  by the Sobolev embedding that
$\partial_{x}u^\e\in\left[L^{p}((0,T)\times\R)\right]^M$ for all
$1<p\le\infty$.

\noindent \underline{{\bf Step 2 (Recurrence)}:}\\

\noindent Now, we use the same steps, we can prove 	
by recurrence that for all $m\in \N$ if, 

$$(H_m)\;\;\left|\begin{array}{ll}
&\partial_{x}u^\e \in\left[L^{\infty}((0,T)\times\R)\right]^M,\\
\\
&\partial_{x}u^\e,\;\;F[u^\e]\in \left[W^{m,p}((0,T)\times\R)\right]^M
\;\;\;\mbox{for all $1<p\le 4$},\end{array}\right.$$

\noindent then  
$$(H_m)\Rightarrow (H_{m+1}).$$

\noindent Indeed, as in  (\ref{EM:etap1}) we can deduce here that 

\begin{equation}\label{EM:etap11}\partial_t u^\e \;\;\;\mbox{and}\;\;\;
  \partial_{xx} u^\e \in
  \left[W^{m,p}((0,T)\times\R)\right]^M,\;\;\mbox{for all $1 <p \le 4$},
\end{equation}

\noindent and From (\ref{EM:D1}), because $\partial_{x}u^\e
\in\left[L^{\infty}((0,T)\times\R)\right]^M$, we can obtain here that

$$\partial_x F[u^\e]\in
\left[W^{m,p}((0,T)\times\R)\right]^M\;\;\;\mbox{for all $1<p\le 4$}.$$

\noindent Which proves that, as in (\ref{EM:etap2}) that 

\begin{equation}\label{EM:etap21}\partial_{tx} u^\e \;\;\;\mbox{and}\;\;\; 
\partial_{xxx} u^\e \in \left[
 W^{m,p}((0,T)\times\R)\right]^M,\;\;\mbox{for all  $1 <p \le
 4$},\end{equation}

\noindent and From (\ref{EM:D2}), we deduce that 

$$\partial_t F[u^\e]\in
\left[W^{m,p}((0,T)\times\R)\right]^M\;\;\;\mbox{for all $1<p\le 4$},$$

\noindent and then

$$\partial_{x}u^\e,\;\;F[u^\e]\in
\left[W^{m+1,p}((0,T)\times\R)\right]^M\;\;\;\mbox{for all $1<p\le 4$},$$

\noindent Which proves by the Sobolev embedding the results. $\hfill\Box$

\begin{lem}\label{EM:lem:e0}{\bf (Classical Maximum Principle)}\\
 Let $T>0$. For all initial data  $u_0^\e\in [X(\R)]^M$, where
 $\partial_x u_0^\e\in [W^{m,p}(\R)]^M$ for all $m\in\N$, $1\le p\le +\infty$, and satisfying $(H3)$.\\

\noindent If $u^{\e}$ is a solution of the system 
(\ref{EM:burgersapp})-(\ref{EM:initialapp}), such that $u^{\e}\in\left[C([0,T); X(\R))\right]^M$ and
 $\partial_x u^{\e}\in [L^{\infty}((0,T);L^1(\R))]^M$, then we have for
 $i=1,\dots,M$,  $\partial_{x} u^{\e,i}\ge 0$   on $(0,T)\times
   \R$.
\end{lem}

\noindent{\bf Proof of Lemma \ref{EM:lem:e0}}\\
\noindent We first derive with respect to $x$ the system
(\ref{EM:burgersapp})-(\ref{EM:initialapp}), and get for
$w^{\e}=(w^{\e,i})_{i=1,\dots,M}= \partial_x u^\e$

$$\partial_t w^\e -\e\partial_{xx}w^\e+ a(u^{\e}) \diamond \partial_{x}w^\e+ Da(u)w^{\e}
\diamond w^{\e}=0.$$

\noindent  Since $u^{\e}\in \left[C^{\infty}([0,T)\times \R)\right]^M$,
we see, for $i=1,\dots,M$, that $w^{\e,i}$ is smooth and satisfies
$w^{\e,i}(0,x)=\partial_x u_0^{\e,i}\ge 0$. From the classical maximum
principle we deduce that $w^{\e,i} \ge 0$ on $[0,T)\times\R$.  $\hfill\Box$

\begin{rem}\label{EM:boru_x}{\bf ($L^{1}$ uniform estimate on $\partial_x
    u^\e$)}\\
Because  $\partial_x u^{\e,i}\ge 0$, for  $i=1,\dots,M$, we deduce from Lemma \ref{EM:Ldeu} that:

\begin{equation}\label{EM:L_1}\left\|\partial_x u^{\varepsilon}\right\|_{\left[L^{\infty}([0,T);L^1(
    \R))\right]^M} \le 
2\left\|u^{\varepsilon}\right\|_{\left[L^{\infty}([0,T)\times\mathbb R)\right]^M}
\le 2\|u_0^\e\|_{\left[L^{\infty}(\R)\right]^M}.\end{equation}
\end{rem}

\begin{cor}\label{EM:theo:exip_regu}{\bf (global existence of nondecreasing
    smooth solutions)}\\
Let $T>0$. The solution given in Theorem \ref{EM:theo:exip} can be chosen such that  
$u^{\e}=(u^{\e,i})_{i=1,\dots,M}$ smooth, satisfies (\ref{EM:requla}) 
and for each $i=1,\dots,M$, $\partial_{x} u^{\e,i}\ge 0$  on
$(0,T)\times \R$.
 
\end{cor}

\noindent The proof of Corollary \ref{EM:theo:exip_regu} is a consequence of
Theorem \ref{EM:theo:exip} and Lemmata \ref{EM:reg}, \ref{EM:lem:e0} and Remark
\ref{EM:boru_x}.

\section{$\e$-Uniform {\it a priori} estimates }
In this Section, we  show some $\e$-uniform estimates  
on the solutions of the  system
(\ref{EM:burgersapp})-(\ref{EM:initialapp}). These estimates  will be used 
 in Section \ref{EM:preuv}  for the passage to the limit as $\e$ tends to zero.

\begin{lem}\label{EM:Ldeu1}{\bf($L^{\infty}$ bound on $u^\e$ and $L^{1}$
    bound on  $\partial_x u^\e$)}\\
Let $T>0$, $0<\e\le 1$ and function $u_0 \in
\left[L^{\infty}(\R)\right]^M$ satisfying $(H3)$. Then the
solution of the system (\ref{EM:burgersapp})-(\ref{EM:initialapp}) given in
Theorem \ref{EM:theo:exip1} with initial data $u_0^\e=u_0\ast\eta_{\e},$
satisfies the following
$\e$-uniform estimates:\\

\noindent $(E1)$ $\left\|u^\e\right\|_{\left[
    L^{\infty}((0,T)\times\R)\right]^M} \le
\left\|u_0\right\|_{\left[L^{\infty}(\R)\right]^M},$\\

\noindent $(E2)$
$\left\|\partial_x{u}^{\e}\right\|_{\left[L^{\infty}((0,T),L^1(\R))\right]^M}
\le 2\left\|u_0\right\|_{\left[L^{\infty}(\R)\right]^M}
,$\\
\end{lem}

\noindent {\bf Proof of Lemma \ref{EM:Ldeu1}:}\\
\noindent First, we remark that if $\partial_x u_0\ge 0$, then  $\partial_x
u_0^\e = (\partial_x u_0)\ast\eta_{\e}(x)\ge 0$ (because  $\eta$ is
positive). The fact that $u_0 \in \left[L^{\infty}(\R)\right]^M$ and
$\partial_x u_0\ge 0$, we obtain that $\partial_x u_0
\in \left[L^{1}(\R)\right]^M$. 

\noindent By classical properties of the mollifier
$(\eta_{\e})_{\e}$ we know that if  $u_0\in \left[L^{\infty}(\R)\right]^M$
and  $\partial_x u_0 \in\left[L^{1}(\R)\right]^M$  we have $u_0^\e\in
\left[X(\R)\right]^M$ and $\partial_x u_0^\e\in [W^{m,p}(\R)]^M$ for all
$m\in\N$, $1\le p\le +\infty$.\\

\noindent Now, we use  Lemma \ref{EM:Ldeu}
and Remark \ref{EM:boru_x}, we deduce by the classical properties of the
mollifier $(E1)$ and $(E2)$.\\

\noindent Before going into the proof of the gradient entropy inequality
defined in (\ref{EM:eq:entropie}), we announce the  main idea of this new
gradient entropy estimate. Now, let us set for $w\ge 0$ the entropy function
$$\bar{f}(w)=w\ln w.$$
For any {\it non-negative} test function $\varphi \in C^1_c(\R\times
[0,+\infty))$, let us define the following {\it ``gradient entropy''} with
$w^i:=\partial_x u^i$:
$$\displaystyle{\bar{N}(t)}= \int_{\R}  \varphi \left(\sum_{i=1,...,M}
\bar{f}(w^i)\right)\ dx.$$
It is very natural to introduce such quantity $\bar{N}(t)$ which in
the case $\varphi\equiv 1$, appears to be nothing else than the total entropy of the
system of $M$ type of particles of non-negative densities $w^i$.
Then  it is formally possible to deduce from (\ref{EM:burger}) the
equality in the following new {\it gradient entropy inequality} for all $t\ge 0$\\
\begin{equation}\label{EM:eq:entropie}
\displaystyle{\frac{d \bar{N}}{dt}(t) + \int_{\R}\varphi \left(\sum_{i,j=1,...,M}
  a^i_{,j}w^iw^j\right)\ dx \le R(t)} \quad \quad \mbox{for}\quad  t\ge 0,
\end{equation}
with the rest
$$R(t)= \displaystyle{\int_{\R }  \left\{(\partial_t\varphi)
\left(\sum_{i=1,...,M}\bar{f}(w^i)\right)+(\partial_x\varphi)
\left(\sum_{i=1,...,M} a^i
 \bar{f}(w^i)\right)\right\}\ dx,}
$$
where we only show the dependence on $t$ in the integrals.
We remark in particular that this rest is formally equal to zero if $\varphi\equiv
1$.\\

\noindent To guarantee the existence of continuous solutions, we assumed in $(H2)$ a sign on
the left hand side of inequality (\ref{EM:eq:entropie}).

\noindent For we return this previous calculate more rigorous, we prove 	
actually the following gradient entropy inequality

\begin{pro}\label{EM:lemme:entro}{\bf(Gradient entropy inequality)}\\ 
Let $T>0$, $0<\e\le 1$ and function $u_0 \in
\left[L^{\infty}(\R)\right]^M$ satisfying $(H3)$. We consider the
solution $u^\e$ of the system (\ref{EM:burgersapp})-(\ref{EM:initialapp}) given in
Theorem \ref{EM:theo:exip1} with initial data
$u_0^\e=u_0\ast\eta_{\e},$. Then, there exists a constant $C(T,M,M_1,
\|u_0\|_{[L^{\infty}(\R)]^M},\|\partial_x u_0\|_{[L\log L
  (\R)]^M}$ such that 

\begin{equation}\label{EM:keyest}
N(t)+\displaystyle{\int_{0}^t\int_{\R}\sum_{i,j=1,\dots,M}a^i_{,j}(u^\e)w^{\e,i}w^{\e,j}}
\le C,\;\;\;\mbox{with}\;\;\; N(t)=\displaystyle{\int_{\R}\sum_{i=1,\dots,M} f(w^{\e,i})dx}.
\end{equation}

\noindent where  $w^{\e}=(w^{\e,i})_{i=1,\dots,M}=\partial_x u^{\e}$ and 
 $f$ is defined in (\ref{EM:f}).

\end{pro}

\noindent  For the proof of Proposition \ref{EM:lemme:entro} we need  the following Lemma:

\begin{lem}{\bf($L\log L$ Estimate)}\label{EM:e(0)}\\
Let $(\eta_\e)_{\e}$ be a non-negative mollifier, $f$ is the function
defined in (\ref{EM:f}) and $h\in L^1(\R)$ is a non-negative function. Then

\noindent i) $\displaystyle{\int_{\R}}f(h)<+\infty$ if and only if
$h\in L\log L(\R)$.\\

\noindent ii) If $h\in L\log L(\R)$ the function
$h_{\e}=h\ast\eta_{\e}\in L\log L(\R)$ satisfies

$$\|h-h_{\e}\|_{L\log L(\R)} \rightarrow 0 \qquad\mbox{as}\qquad
\e\rightarrow 0.$$ 

\end{lem}

\noindent  The proof of (i) is trivial, for the proof of (ii) see R. A. Adams
\cite[Th 8.20]{Adams} for the proof of this Lemma.

\noindent {\bf Proof of Proposition \ref{EM:lemme:entro}:}\\
\noindent Remark first that the quantity $N(t)$ is well-defined because
$w^{\e}\in \left[L^{\infty}((0,T);L^1(\R))\right]^M \cap
\left[L^{\infty}((0,T);L^8(\R))\right]^M$ (by Theorem \ref{EM:theo:exip}
and Corollary \ref{EM:theo:exip_regu}) and we have the general
inequality $\frac{-1}{e}\le w\log w \le w^2$ for all $w\ge 0$.\\

\noindent From Theorem \ref{EM:theo:exip_regu} we know that $w^{\e,i}$ and
smooth non-negative function. Now, we derive $N(t)$ with respect to $t$,
this is well-defined because for $i=1,\dots,M$, we have $\d{\left|\int_{w^{\e,i}\ge \frac {1}{e}}\right|\le
e\|w^{\e,i}\|_{L^{\infty}((0,T);L^1(\R))}}$ and for all $m\in \N$,
$w^{\e,i}\in W^{m,\infty}((0,T)\times\R)$ (see (\ref{EM:requla})).\\

\noindent Finally, we get that,

$$\begin{array}{lll}
\displaystyle{\frac{d}{dt}N(t)}
&=\displaystyle{\int_{\R}\sum_{i=1,\dots,M}
  f'(w^{\e,i})(\partial_t w^{\e,i})},\\\\
&=\displaystyle{\int_{\R}\sum_{i=1,\dots,M}
f'(w^{\e,i})\partial_{x}\left(-a^i(u^{\e}) w^{\e,i}
+\varepsilon\partial_{x}
w^{\varepsilon,i}\right)},\\\\
&=\overbrace{
  \mathstrut\displaystyle{\int_{\R}}\sum_{i=1,\dots,M} a^i(u^{\e})
w^{\e,i}f''(w^{\e,i})\partial_{x} w^{\e,i}}^{J_1}
\;\overbrace{
  \mathstrut-\displaystyle{\varepsilon\int_{\R}\sum_{i=1,\dots,M}\left(\partial_x
w^{\e,i}\right)^2f''(w^{\e,i})}}^{J_2}
\end{array}$$

\noindent But, it is easy to check that 
$$f'(x)=\left\{\begin{array}{ll}
\ln(x)+1 & \quad \mbox{if}\quad x \ge 1/e,\\
0 & \quad \mbox{if}\quad 0\le x \le 1/e,\\
\end{array}\right. \;\;\; \mbox{and}\;\;\; f''(x)=\left\{\begin{array}{ll}
\frac{1}{x} & \quad \mbox{if}\quad x \ge 1/e,\\
0 & \quad \mbox{if}\quad 0\le x \le 1/e.\\
\end{array}\right.$$

\noindent This proves that $J_2\le 0$. To control  $J_1$, we rewrite it
under the following form 

$$J_1= \displaystyle{\int_{\R}}\sum_{i=1,\dots,M} a^i(u^{\e})
g'(w^{\e,i})\partial_{x} w^{\e,i},$$

\noindent where 

$$g(x)=\left\{\begin{array}{ll}
x-\frac {1}{e} & \quad \mbox{if}\quad x \ge 1/e,\\
0 & \quad \mbox{if}\quad 0\le x \le 1/e,\\
\end{array}\right.$$

\noindent Then,  we deduce that 

$$\begin{array}{lll}J_1
&=\displaystyle{\int_{\R}}\sum_{i=1,\dots,M} a^i(u^{\e})
\partial_{x}(g(w^{\e,i}))\\ 
&=-\displaystyle{\int_{\R}}\sum_{i,j=1,\dots,M} a^i_{,j}(u^{\e})w^{\e,j}
g(w^{\e,i}),\\
&=\overbrace{
  \mathstrut-\displaystyle{\int_{\R}}\sum_{i,j=1,\dots,M}
  a^i_{,j}(u^{\e})w^{\e,j}w^{\e,i}}^{J_{11}}
\;\overbrace{
  \mathstrut-\displaystyle{\int_{\R}}\sum_{i,j=1,\dots,M}
a^i_{,j}(u^{\e})w^{\e,j}(g(w^{\e,i})-w^{\e,i})}^{J_{12}},
\end{array}$$

\noindent From $(H2)$, we know that $J_{11}\le 0$. We use the fact that
$|g(x)-x|\le \frac {1}{e}$ for all $x\ge 0$ and $(H1)$, to deduce that

$$\begin{array}{lll}|J_{12}|
&\le\frac {1}{e}
M^2M_1\left\|w^{\e,i}
\right\|_{\left[L^{\infty}((0,T),L^1(\R))\right]^M}\\
\\
&\le\frac {2}{e}
M^2M_1\|u_0\|_{[L^{\infty}(\R)]^M}
\end{array}$$

\noindent where we have use Lemma \ref{EM:Ldeu1} $(E2)$ in the last
line. Finally, we deduce that, there exists a positive constant
$C(\|u_0\|_{[L^{\infty}(\R)]^M},M_1,M)$ independent of $\e$ such that
 
$$\begin{array}{lll}\displaystyle{\frac{d}{dt}N (t)}
&\le J_{11}+J_{12}+J_{2}\\
&\le J_{11}+C
.\end{array}$$

\noindent Integrating in time we get by Lemma \ref{EM:e(0)}, there exists a
another positive constant $C(T,M,M_1,
\|u_0\|_{[L^{\infty}(\R)]^M},\|\partial_x u_0\|_{[L\log L
  (\R)]^M})$ independent of $\e$ such that  

$$N(t)+\displaystyle{\int_{0}^t\int_{\R}}\sum_{i,j=1,\dots,M}
  a^i_{,j}(u^{\e})w^{\e,j}w^{\e,i}\le CT+N(0)\le C.$$

$\hfill\Box$

\begin{lem}{\bf($W^{-1,1}$ estimate on the time derivatives of the
    solutions)}\label{EM:lem:etem}\\
Let $T>0$, $0<\e\le 1$ and function $u_0 \in
\left[L^{\infty}(\R)\right]^M$ satisfying $(H3)$. Then the
solution of the system (\ref{EM:burgersapp})-(\ref{EM:initialapp}) given in
Theorem \ref{EM:theo:exip1} with initial data $u_0^\e=u_0\ast\eta_{\e},$
satisfies the following
$\e$-uniform estimates:\\

$$\left\|\partial_t u^{\e} \right\|_{ \left[L^{2}((0,T); W^{-1,1}(\R))\right]^M}\leq
C\left(1+\|u_0\|_{\left[L^{\infty}(\R)\right]^M}^2\right).$$

\noindent where $W^{-1,1}(\R)$ is  the dual of the space $W^{1,\infty}(\R).$

\end{lem}
\noindent {\bf Proof of Lemma \ref{EM:lem:etem}:}\\
\noindent The idea to  bound
$\partial_t u^{\e}$ is simply to use  the available 
bounds on the right hand side of the equation (\ref{EM:burgersapp}).

\noindent  We will give a proof by duality. We multiply the equation (\ref{EM:burgersapp}) by
$\phi\in \left[L^{2}((0,T), W^{1,\infty}(\R))\right]^M$ and integrate on
$(0,T)\times \R$,  which gives

$$\displaystyle{
\int_{(0,T)\times\R}\phi\;\partial_t u^\e =
\overbrace{
  \mathstrut\e\int_{(0,T)\times\R}\phi\;\partial_{xx}^2u^\e}^{I_1}
\;\overbrace{ \mathstrut-\int_{(0,T)\times\R}\phi\;a(u^\e) \diamond \partial_x u^\e}^{I_2}}.
$$
\noindent We integrate by parts the term $I_1$, and obtain that for
$0< \e\le 1$:
\begin{equation}\begin{array}{ll}\label{EM:I_1}\displaystyle{|I_1|\le
\left|\int_{(0,T)\times\R}\partial_{x}\phi\partial_{x}u^\e\right|}
&\le
\displaystyle{T\|\partial_{x}\phi\|_{\left [L^{2}((0,T), L^{\infty}(\R))\right]^M}
    \|\partial_{x}u^\e\|_{\left [L^{2}((0,T), L^{1}(\R))\right]^M}},\\
\\
&\displaystyle{\le
  2T^{\frac 32}\|\phi\|_{\left[L^{2}((0,T), W^{1,\infty}(\R))\right]^M}
\|u_0\|_{\left [L^{\infty}(\R)\right]^M}},
\end{array}\end{equation}
\noindent here, we have  used the inequality 
\begin{equation}\label{EM:L_2}\left\|\partial_x u^{\varepsilon}\right\|_{\left[L^{2}([0,T);L^1(
    \R))\right]^M} \le  2 T^{\frac
  12}\|u_0\|_{\left[L^{\infty}(\R)\right]^M},
\end{equation}

\noindent which follows from  estimate (\ref{EM:L_1}) for bounded and
nondecreasing function $u^\e$.
Similarly, for the term $I_2$, we have:  

\begin{equation}\begin{array}{ll}\label{EM:I_2}|I_2|
&\le\displaystyle{ M_0\|u\|_{\left
      [L^{\infty}((0,T)\times\R)\right]^M}
\|\phi\|_{\left
      [L^{2}((0,T), L^{\infty}(\R))\right]^M}
\|\partial_{x}u^\e\|_{\left [L^{2}((0,T), L^{1}(\R))\right]^M}},\\
\\
&\le \displaystyle{2T^{\frac 12}M_0\|u_0\|_{\left
      [L^{\infty}(\R)\right]^M}^2\|\phi\|_{\left
[L^{2}((0,T), W^{1,\infty}(\R))\right]^M}}.
\end{array}\end{equation}

\noindent Finally,  collecting (\ref{EM:I_1}) and (\ref{EM:I_2}), we get that
there exists a constant $C=C(T,M_0)$ independent of
$0<\e\le 1$ such that:
$$\displaystyle{\left|
\int_{(0,T)\times\R}\phi\partial_t u^\e \right|
\le C\left(1+\|u_0\|_{\left
      [L^{\infty}(\R)\right]^M}^2\right)\|\phi\|_{\left
[L^{2}((0,T), W^{1,\infty}(\R))\right]^M}}
$$
which gives the announced result where we use that $L^{2}((0,T),
W^{-1,1}(\R))$ is the dual of $L^{2}((0,T),
W^{1,\infty}(\R))$ (see   Cazenave and  Haraux
  \cite[Th 1.4.19, Page 17]{Cazen}). $\hfill\Box$

\begin{cor}\label{EM:born}{\bf ($\e$-Uniform estimates)}\\
Let $T>0$, $0<\e\le 1$ and function $u_0 \in
\left[L^{\infty}(\R)\right]^M$ satisfying $(H1)$ and $(H2)$. Then the
solution of the system (\ref{EM:burgersapp})-(\ref{EM:initialapp}) given in
Theorem \ref{EM:theo:exip1} with initial data $u_0^\e=u_0\ast\eta_{\e},$
satisfies the following
$\e$-uniform estimates:\\

$$\left\|\partial_ x{u}^{\e}\right\|_{\left[ L^{\infty}((0,T);  L\log L(\R))\right]^M}
+\left\|u^\e\right\|_{\left[ L^{\infty}((0,T)\times \R)\right]^M}
+\left\|\partial_t{u}^{\e}\right\|_{\left[ L^{2}((0,T); W^{-1,1}(\R))\right]^M}
\le C.$$

\noindent where  $C=C(T,M,M_0,M_1
\left\|u_0\right\|_{\left[L^{\infty}(\R)\right]^M},\left\|
\partial_x u_0\right\|_{\left[L\log L(\R)\right]^M})$.

\end{cor}

\noindent We can easily prove this Corollary collecting Lemmata
\ref{EM:Ldeu1}, \ref{EM:lem:etem} and \ref{EM:e(0)} and Proposition \ref{EM:lemme:entro}.
 \section{Passage to the limit and the proof of Theorem \ref{EM:th1}}\label{EM:preuv}

In this section, we prove that the system
 (\ref{EM:burger})-(\ref{EM:initialdata}) admits solutions $u$ in the
distributional sense.  They are the limits  of $u^{\e}$ given by
Theorem \ref{EM:theo:exip1} when  $\e\rightarrow
0$. To do this, we
will justify the passage  to the limit as $\e$ tends to $0$ in the
system (\ref{EM:burgersapp})-(\ref{EM:initialapp}) by using some 
compactness  tools that are presented in a first Subsection. 
\subsection{Preliminary results}
First,  for all $I$  open interval of $\R$, we denote by 
$$L\log L(I)==\left\{\mbox{$f\in L^1(I)$ such that
  $\displaystyle{\int_{I}|f|\ln\left(1+|f|\right)<+\infty}$}
 \right\}.$$

\begin{lem}\label{EM:lem:comp}{\bf (Compact embedding)}\\
\noindent Let $I$ an open and bounded interval of $\R$. If we denote
by
$$W^{1, L \log L}(I)=\{\mbox{$u\in L^{1}(I)$ such that
  $\partial_x u\in L\log L(I)$}\}.$$

\noindent Then the following injection:

$$W^{1, L \log L}(I)\hookrightarrow C(I),$$
\noindent is compact.
\end{lem}
\noindent For the proof of this Lemma see R. A. Adams \cite[Th
8.32]{Adams}.

\begin{lem}\label{EM:simo}{\bf (Simon's Lemma)}\\
\noindent Let $X$, $B$, $Y$ be three  Banach spaces, such that

$$\mbox{$X\hookrightarrow B$ with compact embedding and $B\hookrightarrow Y$ with continuous
 embedding}.$$ 

\noindent Let $T>0$. If $(u^\e)_\e$ is a sequence such that,

$$\|u^\e\|_{L^{\infty}((0,T); X)}+\|u^\e\|_{L^{\infty}((0,T); B)}+
\left\|\partial_t u^\e\right\|_{L^q((0,T); Y)}\le C,
$$

\noindent  where $q>1$ and $C$ is a constant independent of $\e$, 
then $(u^\e)_\e$ is relatively compact in $C((0,T); B)$.

\end{lem}
\noindent For the proof, see J. Simon \cite[Corollary 4, Page 85]{SI87}.\\

\noindent In order to show the existence of solution system (\ref{EM:burger}) in
Subsection \ref{EM:preuvf}, we will apply this lemma to each scalar component in the particular case
where  $X=W^{1,\log}(I)$, $B= L^{\infty}(I)$ and
$Y=W^{-1,1}(I):=(W^{1,\infty}(I))'$.\\

\noindent We denote by $K_{exp}(I)$ the class of all measurable function $u$,
defined on $I$, for which,

$$\displaystyle{\int_{I}\left(e^{|u|}-1\right)<+\infty}.$$
 
\noindent The space  $EXP(I)$ is defined to be the linear hull of
$K_{exp}(I)$. This space is supplemented with the following Luxemburg
norm (see Adams \cite[(13), Page
234]{Adams} ):
 
$$\|u\|_{EXP(I)}=\inf\left\{\lambda>0:\displaystyle{\int_{I}
\left(e^{\frac{|u|}{\lambda}}-1\right)\le 1}\right\},$$

\noindent Let us recall some useful properties of this space.

\begin{lem}\label{EM:weak}{\bf (Weak star topology in $L\log L$)}\\
Let $E_{exp}(I)$ be the closure in $EXP(I)$ of the
space of functions  bounded on $I$. Then $E_{exp}(I)$  is a separable
Banach  space which verifies, \\

\noindent i) \centerline{$\mbox{$L\log
L(I)$ is the dual space of $E_{exp}(I)$.}$} \\

\noindent  ii) \centerline{$\mbox{$L^{\infty}(I) 
\hookrightarrow E_{exp}(I)$.}$}

\end{lem}
\noindent For the proof, see Adams \cite[Th 8.16, 8.18,
8.20]{Adams}.

\begin{lem}\label{EM:hold}{\bf(Generalized Hölder inequality, Adams 
    \cite[8.11, Page 234]{Adams})}\\
Let $f\in EXP (I)$ and $g\in L\log L(I)$. Then $fg\in L^1(I)$, with 

$$\|fg\|_{L^1(I)}\le 2\|f\|_{EXP(I)}\|g\|_{L\log L(I)}.$$
\end{lem}

 \noindent The following Lemma, we allow to define later the restriction of a function 
$f\in W^{-1,1}(\R)$ on all open interval $I$ of $\R$.

\begin{lem}{\bf (Extension)}\\
For all open interval 
$I$ of $\R$, there exists a linear and continuous operator of extension $P:
W^{1,\infty}(I)\rightarrow W^{1,\infty}(\R)$ such that\\

\noindent i) $Pu_{|_I}=u$ for  $u\in W^{1,\infty}(I)$.\\

\noindent ii) $\|Pu\|_{W^{1,\infty}(\R)}\le\|u\|_{W^{1,\infty}(I)}$ for 
$u\in W^{1,\infty}(I)$.
\end{lem}
\noindent for the proof of this Lemma see for instance  Brezis
\cite[Th.8.5]{Bre}.\\

\noindent Thanks this Lemma, we can notice that, if $f\in
W^{-1,1}(\R)$, where $W^{-1,1}(\R):=(W^{1,\infty}(\R))'$, we can define,  for all
open interval $I$ of $\R$, the function $f_{|_I}$ as the
following

$$<f_{|_I},h>_{W^{-1,1}(I),W^{1,\infty}(I)}=<f,
Ph>_{W^{-1,1}(\R),W^{1,\infty}(\R)}.$$

\noindent for all $h\in W^{1,\infty}(I)$.


\subsection{Proof of  Theorem \ref{EM:th1}}\label{EM:preuvf}

\noindent \underline{ {\bf Step 1 (Existence)}:}\\

\noindent  First, by Corollary \ref{EM:born}  we know that for any $T>0$, the
solutions $u^{\e}$ of the system
(\ref{EM:burgersapp})-(\ref{EM:initialapp})  obtained with
the help of Theorem  \ref{EM:theo:exip1},  are $\e$-uniformly bounded in
$\left[L^{\infty}((0,T)\times\R)\right]^M$. Hence,  as $\e$
goes to zero, we  can extract a subsequence   still denoted
 by $u^{\e}$, that converges weakly-$\star$  in
$\left[L^{\infty}((0,T)\times\R)\right]^M$ to some limit  $u$. 
Then we want to  show that $u$ is a solution of the
system (\ref{EM:burger})-(\ref{EM:initialdata}). Indeed, since the passage to the
limit in the linear terms is trivial in $\left[\D'((0,T)\times\R)\right]^M$, it
suffices to pass to the limit in the non-linear term,

$$a(u^\e)\diamond \partial_x u^\e.$$

\noindent According to Corollary \ref{EM:born} we know that for all open and
bounded interval $I$ of $\R$ there exists a constant $C$  independent on $\e$
such that:

$$\left\|{u}^{\e}\right\|_{\left[ L^{\infty}((0,T); W^{1, L\log L}(I))\right]^M}
+\left\|u^\e\right\|_{\left[ L^{\infty}((0,T)\times I)\right]^M}
+\left\|\partial_t{u}^{\e}\right\|_{\left[ L^{2}((0,T); W^{-1,1}(I))\right]^M}
\le C.$$

\noindent From the compactness of $W^{1,L\log L}(I)\hookrightarrow
L^{\infty}(I)$ (see Lemma \ref{EM:weak} (i)),
we can apply Simon's Lemma (i.e. Lemma \ref{EM:simo}), 
with $X=\left[W^{1,L\log L}(I)\right]^M$, $B=\left[L^{\infty}(I)\right]^M$
and $Y=\left[W^{-1,1}(I)\right]^M$, which shows that

\begin{equation}\label{EM:Linf}\mbox {$u^\e$ is relatively  compact in
 in $\left[L^{\infty}((0,T)\times I)\right]^M \hookrightarrow
\left[L^{1}((0,T); L^{\infty}(I))\right]^M.$}\end{equation}

\noindent  Then form continuous
injection of $L^{\infty}(I) \hookrightarrow E_{exp}(I)$  (see Lemma
\ref{EM:weak} (ii)), we deduce that,

\begin{equation}\label{EM:compact}\mbox{$u^\e$ is relatively  compact in $\left[L^{1}((0,T);
  E_{exp}(\Omega))\right]^M$.}\end{equation}

\noindent On the other hand,
by Corollary \ref{EM:born}, we notice that
$\partial_x u^\e$ is $\e$-uniformly bounded in
  $\left[L^{\infty}((0,T); L\log L(I))\right]^M$. Moreover, the space
  $\left[L^{\infty}((0,T); L\log L(I))\right]^M$ is the dual space of $\left[L^{1}((0,T);
  E_{exp}(I))\right]^M$, because $L\log L(I)$ is the dual space of
$E_{exp}(I)$ (see Lemma \ref{EM:weak} (ii) and Cazenave, Haraux
\cite[Th 1.4.19, Page 17]{Cazen}). Then, up to a subsequence

\begin{equation}\label{EM:compact1}\mbox{$\partial_x u^\e \rightarrow \partial_x u$  
weakly-$\star$ in $\left[L^{\infty}((0,T); L\log L(I))\right]^M$ 
.}\end{equation}

\noindent Form (\ref{EM:compact}) and (\ref{EM:compact1}), we see that we can
pass to the limit in the non-linear term in the sense 

$$\left[L^{1}((0,T); E_{exp}(I))\right]^M-strong\;
\times\;\left[ L^{\infty}((0,T); L\log
L(I))\right]^M-weak-\star.$$

\noindent Because this is true for any bounded open interval $I$ and for
any $T>0$, we deduce that,

$$a(u^\e)\diamond \partial_x u^\e  \rightarrow a(u) \diamond \partial_x u\;\;\; \mbox{in}\;\;
\D'((0,T)\times \R)$$

\noindent  Consequently, we can pass to the limit in (\ref{EM:burgersapp}) and get that,

$$\partial_t{u}+a(u)\diamond \partial_x u =0\;\;\; \mbox{in}\;\;
\D'((0,T)\times \R).$$

\noindent  This solution $u$ is also satisfy the
following estimates (see for instance
 Brezis \cite[Prop. 3.12]{Bre}):\\

\noindent $(E1')$ $\left\|\partial_x{u}\right\|_{\left[ L^{\infty}((0,T);L\log
    L(\R))\right]^M}\le \liminf
\left\|\partial_x{u}^{\e}\right\|_{\left[ L^{\infty}((0,T);L\log
L(\R))\right]^M} \le C ,$\\

\noindent $(E2')$ $\|u\|_{\left[L^{\infty}((0,T)\times \R)\right]^M}\le 
\liminf  \left\|u^\e\right\|_{\left[
     L^{\infty}((0,T)\times\R)\right]^M} \le
 \left\|u_0\right\|_{\left[L^{\infty}(\R)\right]^M},$\\

\noindent   At this stage we remark that, thanks to these two estimates
we obtain that   $(a(u)\diamond \partial_x u)\in
\left[L^{\infty}((0,T);L\log L(\R))\right]^M$, which gives, since 
$\partial_t{u}=-a(u)\diamond \partial_x u$, that  $\partial_t {u}\in\left[
  L^{\infty}((0,T);L\log L(\R))\right]^M$, and then
  $ u \in \left[C([0,T);L\log L(\R))\right]^M$.\\

\noindent \underline{{\bf Step 2 (The initial conditions)}:}\\

\noindent It remains to prove that the initial conditions (\ref{EM:initialdata}) coincides
 with $u(\cdot,0)$. Indeed, by Corollary \ref{EM:born}, we see that, for all
 open bounded interval  $I$ of $\R$, $u^{\e}$ is  $\e$-uniformly bounded in 

$$\left[W^{1,2}((0,T); W^{-1,1}(I))\right]^M \hookrightarrow\left[C^{\frac 12}([0,T);
W^{-1,1}(I))\right]^M,$$ 

\noindent where $W^{-1,1}(I)$ is the dual of $W^{1,\infty}(I)$.
 It follows that, there exists a
constant $C$  independent on $\e$, such that, for all $t,s\in [0,T)$:
$$\|u^\e(t)-u^\e(s)\|_{\left[W^{-1,1}(I)\right]^M}\le C
|t-s|^{\frac 12}.$$

\noindent In particular if we set $s=0$, we have:
\begin{equation}\label{EM:hol}\|u^{\e}(t)- u^{\e}_0\|_{\left[W^{-1,1}(I)\right]^M}\le C
t^{\frac 12}.\end{equation}
\noindent Now we pass to the limit in (\ref{EM:hol}). Indeed, 
the functions $u^{\e}$ and $u^{\e}_0$ are $\e$-uniformly 
bounded in $\left[W^{1,2}((0,T); W^{-1,1}(I))\right]^M$ and  $\left[W^{-1,1}(I)\right]^M$
respectively. Moreover we  know that
$u^{\e}-u^{\e}_0$ converges weakly-$\star$ in
$\left[L^{\infty}( (0,T)\times I)\right]^M$ to $u-u_0$.\\ 

\noindent Therefore, we
can extract a subsequence still denoted by $u^{\e}-u^{\e}_0$, that
weakly-$\star$  converges in $\left[W^{1,2}((0,T);  W^{-1,1}(I))\right]^M$ to
$u-u_0$. In particular this subsequence 	
 converges, for all $t\in (0,T)$, weakly-$\star$ in $\left[L^{\infty}((0,t);
W^{-1,1}(I))\right]^M$, and consequently it verifies (see for instance
 Brezis \cite[Prop. 3.12]{Bre}),

$$\|u-u_0\|_{\left[L^{\infty}((0,t);W^{-1,1}(I))\right]^M}\le
\liminf  
\|u^\e-u^\e_0\|_{\left[L^{\infty}((0,t);W^{-1,1}(I))\right]^M}\le
Ct^{\frac 12}.$$

\noindent From (\ref{EM:hol}) we deduce that
$$\|u(t)-u_0\|_{\left[W^{-1,1}(I)\right]^M}\le
Ct^{\frac 12},$$
\noindent which proves that  $u(\cdot,0)=u_0$ in
$\left[\D'(\R)\right]^M$.\\

\noindent \underline{{\bf Step 3 (Continuity of solution)}:}\\
Now, we are going to prove the continuity estimate (\ref{EM:contu}). For
all $h>0$ and $(t,x)\in (0,T)\times \R$, we have:

$$\begin{array}{ll}|u(t,x+h)-u(t,x)|
&\le \d{\left|\int_x^{x+h} \partial_x u(t,y)dy\right|}\\
\\
& \le 2 \d{\|1\|_{EXP(x,x+h)}\|\partial_x u\|_{L\log L(x,x+h)}},\\
\\
& \le 2 \d{\frac{1}{\ln(\frac {1}{h}+1)}\|\partial_x u\|_{L^{\infty}((0,T); L\log
  L(\R))}},\\
\\
& \le \d{C\frac{1}{\ln(\frac {1}{h}+1)}},
\end{array}$$

\noindent where we have used in the second line the generalized Hölder
inequality (see Lemma \ref{EM:hold}) and in last line we have used that
$\partial_x u \in L^{\infty}((0,T); L\log L(\R))$. Which proves finally
the continuity in space. Now, we prove the continuity in time, for all
$\delta >0$ and $(t,x)\in (0,T)\times \R$, we have:

$$\begin{array}{llll}\delta|u(t+\delta,x)-u(t,x)|
&=\d{\int_x^{x+\delta}|u(t+\delta,x)-u(t,x)|dy},\\
\\
&\le \overbrace{
  \mathstrut{\d{\int_x^{x+\delta}|u(t+\delta,x)-u(t+\delta,y)|dy}}}^{K_1},\\
&\;\;\;+\overbrace{
  \mathstrut{\d{\int_x^{x+\delta}|u(t+\delta,y)-u(t,y)|dy}}}^{K_2},\\
&\;\;\;+\overbrace{
  \mathstrut{\d{\int_x^{x+\delta}|u(t,y)-u(t,x)|dy}}}^{K_3}.
\end{array}$$

\noindent Similarly, as in the last estimate, we can show that:
 
$$\begin{array}{llll}
K_1+K_3&\le \delta \d{\int_x^{x+\delta}|\partial_x  u(t+\delta,y)|dy},
+\delta \d{\int_x^{x+\delta}|\partial_x  u(t,y)|dy},\\
\\
&\le 4\delta \d{\|1\|_{EXP(x,x+\delta)}\|\partial_x
  u\|_{L^{\infty}((0,T); L\log L(\R))}},\\
\\
&\le \d{C\frac{\delta}{\ln(\frac {1}{\delta}+1)}}.
\end{array}$$

\noindent Now, we use that $u$ is a solution of (\ref{EM:burger}), and
we obtain that:

$$\begin{array}{llll}
K_2
&\le \d{\int_x^{x+\delta}\int_t^{t+\delta}|\partial_t u(s,y)|dy},\\
\\
&\le \d{\int_t^{t+\delta}\int_x^{x+\delta}|a(u(s,y))\diamond \partial_x
  u(s,y)|ds dy},\\
\\
&\le \d{\delta M_0\|u\|_{L^{\infty}((0,T)\times \R )}}
  \d{\|1\|_{EXP(x,x+\delta)}\|\partial_x u\|_{L^{\infty}((0,T); L\log L(\R)}},\\
\\
&\le\d{C\frac{\delta}{\ln(\frac {1}{\delta}+1)},}
\end{array}$$

\noindent  where we have used in  last line that $u\in
L^{\infty}((0,T)\times \R )$, collecting the estimates of 
$K_1$, $K_2$ and $K_3$, we prove that:  

$$|u(t+\delta,x)-u(t,x)|\le \frac{1}{\delta}(K_1+K_2+K_3)\le
\d{C\frac{1}{\ln(\frac {1}{\delta}+1)}},$$
 
\noindent which proves finally the following:

$$|u(t+\delta,x+h)-u(t,x)|\le C\left(
 \displaystyle{\frac{1}{\ln(\frac
    {1}{\delta}+1)}+\frac{1}{\ln(\frac {1}{h}+1)}}\right).$$

$\hfill\Box$


\section{Some remarks on the uniqueness
  }
In this Section we study the uniqueness of solution of the system
(\ref{EM:burger})-(\ref{EM:initialdata}) with
$$\d{a^i(u)=\sum_{j=1,\dots,M}A_{ij}u^j}.$$ 

\noindent  We show some uniqueness
results for some particular matrices with $M\ge2$.

\noindent For the proof of Theorem \ref{EM:unicite1} in Subsection
\ref{EM:uni}, we need to recall in the following Subsection the definition
of viscosity solution and some well-known results in this framework.

\subsection{Some useful results for viscosity solutions}
The notion  of viscosity solutions is quite recente. This concept
has been introduced by  Crandall and Lions \cite{Lio81, CL82} in 1980, to solve 
the first-order Hamilton-Jacobi equations. The theory then extended to
the second order equations by the  work of  Jensen \cite{Jen} and Ishii
\cite{Ishii89}. For good introduction of this theory, we refer to  Barles
\cite{B94} and Bardi, Capuzzo-Dolcetta \cite{BCD97}.

\noindent Now, we recall the definition of viscosity solution for the
following problem for all $0\le\e \le 1$:

\begin{equation}\label{EM:Ham}\partial_t v + H(t,x,v,\partial_{x} v)-\e\partial_{xx}v=0\;\;\;\mbox{with}
\;\;\;x,v\in\R,\; t\in (0,T).\end{equation}

\noindent where  $H :(0,T)\times \R^3\longmapsto \R$ is the Hamiltonian
and is supposed continuous. We will set 
 
$$USC((0,T)\times\R)=\{\mbox{$f$ such that $f$ is upper semicontinuous 
on $(0,T)\times\R$}\},$$
$$LSC((0,T)\times\R)=\{\mbox{$f$  such that $f$ is lower semicontinuous 
  on $(0,T)\times\R$}\}.$$

\begin{defi}\label{EM:defi:salution}{\bf (Viscosity subsolution,
    supersolution and solution)}\\
\noindent A function $v \in USC( (0,T)\times\R)$ is a viscosity subsolution of
(\ref{EM:Ham}) if it satisfies, for every $(t_0,x_0)\in (0,T)\times\R$ and
for every test function $\phi\in C^2((0,T)\times\R)$, that is tangent
from above to $v$ at  $(t_0,x_0)$, the following holds:

$$\partial_t \phi + H(t_0,x_0,v,\partial_{x} \phi)-\e\partial_{xx}\phi \le
0.$$

\noindent A function $v \in LSC( (0,T)\times\R)$ is a viscosity supersolution of
(\ref{EM:Ham}) if it satisfies, for every $(t_0,x_0)\in (0,T)\times\R$ and
for every test function $\phi\in C^2((0,T)\times\R)$, that is tangent
from below to $v$ at  $(t_0,x_0)$, the following holds:
$$\partial_t \phi + H(t_0,x_0,v,\partial_{x} \phi)-\e\partial_{xx}\phi \ge
0.$$

\noindent A function $v$ is a viscosity solution of (\ref{EM:Ham}) if, and
only if, it is a sub and a supersolution of (\ref{EM:Ham}).

\end{defi}

\noindent Let us now recall some well-known results.

\begin{rem}\label{EM:Clas}{\bf (Classical solution-viscosity solution)}\\
If  $v$ is a $C^2$ solution of (\ref{EM:Ham}), then $v$ is a viscosity
solution of (\ref{EM:Ham}).
\end{rem}

\begin{lem}\label{EM:Sta}{\bf (Stability result}, see Barles \cite[Th
  2.3]{B94}{\bf )}\\
We suppose that, for $\e >0$, $v^\e$ is a viscosity solution of 
(\ref{EM:Ham}). If $v^\e\rightarrow v$ uniformly on every compact set  then 
$v$  is a viscosity solution of (\ref{EM:Ham}) with $\e =0$.
 \end{lem}

\begin{lem}\label{EM:Comp}{\bf (Gronwall for viscosity solution)}\\
Let $v$, a locally bounded $USC(0,T)$ function, which is a viscosity subsolution  of the
equation $\d{\frac{d}{dt}v=\a v}$ where $\a\ge 0$. Assume that $v(0)\le v_0$ then
$v\le v_0 \;e^{\a T}$ in $(0,T)$.
\end{lem}

\noindent The proof of this Lemma is a direct application of the
comparison principle, (see Barles \cite[Th 2.4]{B94}).

\begin{rem}{$\;$}\\
From Lemmata \ref{EM:Clas}, \ref{EM:Sta} and  from (\ref{EM:Linf}), we can notice that the solution $u^i$
of our system (\ref{EM:burger}) given in  Theorem \ref{EM:th1} is also a viscosity
solution of (\ref{EM:burger}) (where the $u^j$ for $j\neq i$ are considered
fixed to apply Definition \ref{EM:defi:salution}).\end{rem}

\subsection{Uniqueness results}\label{EM:uni}
In this Subsection we prove Theorem \ref{EM:unicite1}. Before going on, we
recall in the following Remark a well-known uniqueness results and we
recall in Theorem  \ref{EM:unicite} the uniqueness results of
$W^{1,\infty}$ solution of (\ref{EM:burger}).

\begin{rem}{\bf (Uniqueness for  quasi-monotone Hamiltonians)}\\
 If the  elements of the matrix $A$ satisfy:
$$\mbox{$\displaystyle{A_{ii}+\sum_{j\neq i,
  A_{ij}<0}A_{ij}\ge 0}$ \;\;\;for all \;\;\; $i=1,\cdots,M$.}$$

\noindent and  if $\partial_x u^i\ge 0$ for $i=1,\dots,M$, 
then we can easily check that the Hamiltonian
 
$$\displaystyle{H_i(u,\partial_x u^i)=\left(\sum_{j=1,\dots
    M}A_{ij}u^j\right)\partial_x u^i},$$

\noindent is quasi-monotone in the sense of Ishii, Koike
\cite[(A.3)]{IK91}. Then the result of Ishii, Koike \cite[Th.4.7]{IK91}
shows that for any initial condition $u_0\in [L^{\infty}(\R)]^M$
satisfying $(H1)$-$(H2)$,  the system (\ref{EM:burger}) satisfies
the comparison principle which implies the uniqueness of the solution.
\end{rem}

\noindent We have the following result which seems quite standard:

\begin{theo}{\bf (Uniqueness of the $W^{1,\infty}$ solution)}\label{EM:unicite}\\
Let $u_0\in [W^{1,\infty}(\R)]^M$ and $T>0$. Then system
(\ref{EM:burger})-(\ref{EM:initialdata})
 admits a  unique solution in $\left[W^{1,\infty}([0,T)\times\R)\right]^M$.  
\end{theo}
	
\noindent The proof of this Theorem is given in Appendix, because we
have not found any proof of such a result in the literature.\\

\noindent {\bf Proof of Theorem \ref{EM:unicite1}:}\\
\noindent Using Theorem \ref{EM:unicite} with
$\d{a^i(u)=\sum_{j=1,\dots,M}A_{ij}u^j}$, it is enough to show that the
system  (\ref{EM:burger})-(\ref{EM:initialdata}) admits a solution in  
$\left[W^{1,\infty}([0,T)\times\R)\right]^M$. To do that,  it is
enough to prove that the solution  $u^\e$ of the approximated system
obtained in Corollary \ref{EM:born} satisfies that  $\partial_x u^{\e}$
is  bounded in  $\left[L^{\infty}((0,T)\times\R)\right]^M$ uniformly in
$0<\e\le 1$.  Indeed, we then get
the same property for $\partial_x u$, where $u$ is the  limit of $u^\e$
as $\e\rightarrow 0$. Moreover, from  the equation (\ref{EM:burger})
satisfied by $u$ and the fact that 

$$u\in\left[L^{\infty}((0,T)\times \R)\right]^M \quad \mbox{and}\quad
\partial_x u\in\left[L^{\infty}((0,T)\times \R)\right]^M,$$ 

\noindent we deduce that $\partial_t u\in\left[
  L^{\infty}((0,T)\times \R)\right]^M$ which shows that $u\in
\left[W^{1,\infty}([0,T)\times\R)\right]^M$.\\

\noindent To simplify, we denote 

$$w^{\e}=\partial_x u^{\e},$$
 
\noindent and we interest in the 
$$\displaystyle{\max_{x\in \R}  w^{\e,i}(t,x)= m_i(t)}.$$ 

\noindent This maximum is reached at least at some point  $x_i(t)$, because $w^{\e,i}\in
  C^{\infty}((0,T)\times \R)\cap W^{1,p}((0,T)\times \R)$ for all $1<
  p \le +\infty$ (see Lemma \ref{EM:reg}, (\ref{EM:requla})).\\

\noindent In the following we prove in the two cases (i) and (ii) defined in Theorem
\ref{EM:unicite1} that $m_i$, for all $i=1,\dots,M$, is bounded uniformly
in $\e$. First,  deriving with respect
to $x$ the equation (\ref{EM:burgersapp}) 
satisfied by $u^\e \in \left[C^{\infty}((0,T)\times \R)\right]^M$, we
can see that $w^{\e}$ satisfies the following equation

\begin{equation}\label{EM:d_u}\partial_t w^{\e,i} -\e\partial_{xx}w^{\e,i}+
  \sum_{j=1, \dots,M}A_{ij}u^{\e,j} \partial_{x}w^{\e,i}+\sum_{j=1, \dots, M}
A_{ij}w^{\e,j} w^{\e,i}=0.\end{equation}

\noindent Now, we prove that $m_i$ is a viscosity subsolution of the following
  equation,

\begin{equation}\label{EM:m_i}\frac{d}{dt} m_i(t)+\sum_{j=1,\dots,M}A_{ij}w^{\e,j}(t,x_i(t))
w^{\e,i}(t,x_i(t))\le 0.\end{equation}

\noindent Indeed, let $\phi\in C^2(0,T)$ a test function, such that
$\phi\ge m_i$ and $\phi(t_0)=m_i(t_0)$ for some $t_0\in (0,T)$. From the
definition of $m_i$, we can easily check that $\phi\ge w^{\e,i}(t,x)$
and $\phi(t_0)=w^{\e,i}(t_0,x_i(t_0))$. But, the fact that $w^{\e,i}\in
C^{\infty}((0,T)\times \R)$, by Remark \ref{EM:Clas} we know that
$w^{\e,i}$  is a viscosity subsolution of (\ref{EM:d_u}). We apply Definition
\ref{EM:defi:salution}, and the fact that
$\partial_x\phi=\partial_{xx}\phi=0$, we get

$$\frac{d}{dt} \phi(t_0)+\sum_{j=1,\dots,M}A_{ij}w^{\e,j}(t_0,x_i(t_0))
w^{\e,i}(t_0,x_i(t_0))\le 0.$$

\noindent Which proves that $m_i$ is a viscosity subsolution of
(\ref{EM:m_i}).\\

\noindent Two cases may accur:\\

\noindent{\bf i)} Here, we consider the case
where $M\ge 2$ and $A_{ij}\ge 0$ for all $j\ge i$. We see the equation
satisfied by $m_1$, we deduce that satisfies (a viscosity subsolution)

$$\d{\frac{d}{dt} m_1(t)
\le -\sum_{j=1,\dots,M}A_{1j}w^{\e,j}(t,x_1(t))
w^{\e,1}(t,x_1(t))\le 0,}$$

\noindent where we have used the fact that, for $j=1,\dots,M$,
$A_{1j}\ge 0$ and  $w^{\e,j}\ge 0$. This proves by Lemma \ref{EM:Comp} (with
$\a=0$) that,

$$\d{m_1(t)\le m_1(0)=w^{\e,1}(t,x_1(t)) \le 
\|\partial_{x} u^{1}_0\|_{L^{\infty}(\R)}}.$$

\noindent We reason by recurrence: we assume that $m_j \le C$ for all
$j\le i$, where $C$ is a positive constant independent of $\e$, and we
prove that  $m_{i+1}$ is bounded uniformly in $\e$. Indeed, we know that
   
$$\begin{array}{ll}\d{\frac{d}{dt} m_{i+1}(t)}
&\d{\le
-\sum_{j=1,\dots,M}A_{i+1,j}w^{\e,j}(t,x_j(t))
w^{\e,i+1}(t,x_{i+1}(t)),}\\
\\
&\d{\le -\sum_{j<i+1}A_{i+1,j}w^{\e,j}(t,x_j(t))
w^{\e,i+1}(t,x_{i+1}(t))}\\
&\d{\;\;\;\;-\sum_{M\ge j\ge i+1}A_{i+1,j}w^{\e,j}(t,x_j(t))
w^{\e,i+1}(t,x_{i+1}(t))},\end{array}$$

\noindent We use that $A_{i+1,j}\ge 0$, for $M\ge j\ge i+1$, we obtain that

$$\begin{array}{ll}\d{\frac{d}{dt} m_{i+1}(t)}
&\d{\le -\sum_{j<i+1}A_{i+1,j}w^{\e,j}(t,x_j(t))
w^{\e,i+1}(t,x_{i+1}(t))},\\
\\ 
&\d{\le C \left(\sum_{j<i+1}|A_{i+1,j}|\right)m_{i+1}(t)}.
\end{array}$$

\noindent This implies by Lemma \ref{EM:Comp}, with $\a=\d{C
  \left(\sum_{j<i+1}|A_{i+1,j}|\right)}$,   that

$$\begin{array}{ll} m_{i+1}(t)
&\d{\le m_{i+1}(0)e^{\a T}},\\
\\
&\le \d{\|\partial_{x} u^{i+1}_0\|_{L^{\infty}(\R)}e^{\a T}}.\end{array}$$

\noindent Which proves that for all $i=1,\dots,M$, $m_i$ is bounded
uniformly in $\e$.\\

\noindent{\bf ii)} Here, we consider the case
where $M\ge 2$ and $A_{ij}\le 0$ for all $i\neq j$.  Taking the sum over
the index $i$, from (\ref{EM:m_i}) we get that the quantity 
$\displaystyle{m(t)=\sum_{i=1,\dots,M}m_i(t)}$ satisfies (a viscosity
subsolution see Bardi et al. \cite{Bardi})

$$\begin{array}{ll}\d{\frac{d}{dt}m(t)}
&\d{\le -\sum_{i,j=1,\dots,M}A_{ij}w^{\e,j}(t,x_i(t))
w^{\e,i}(t,x_i(t))},\
\\
&\d{\le -\sum_{i,j=1,\dots,M}A_{ij}w^{\e,j}(t,x_j(t))
w^{\e,i}(t,x_i(t))},\
\\
&\le 0.\end{array}$$

\noindent where we have used that the matrix $A$ satisfies $(H2')$ and $w^{\e,i}\ge
0$, for $i=1,\dots,M$. Using Lemma \ref{EM:Comp} with $\a=0$, 
we get 

$$\begin{array}{ll}m(t)
&\d{\le m(0)
=\d{\sum_{i=1,\dots,M}\partial_x u^{\e,i}_0}},\\
\\
&\le \d{\sup_{y\in \R}\sum_{i=1,\dots,M}\partial_x u^{i}_0(y)}.
\end{array}$$

\noindent which proves (\ref{EM:w_infty}). $\hfill\Box$

\section{Application on the dynamics of dislocations densities}\label{EM:subsce:model}
In this Section, we present a model describing the dynamics of
dislocations densities. We refer to \cite{Hirth} for a physical
presentation of dislocations which are (moving) defects in crystals.
Even if the problem is naturally a three-dimensional problem,we will
first assume that the  geometry of the problem is invariant by
translations in the $x_3$-direction. This reduces the problem to the
study of dislocations densities defined on the plane $(x_1,x_2)$ and 
propagation in a given direction $\vec{b}$ belonging to the plane
$(x_1,x_2)$ (which is called the ``Burger's vector'').\\

\noindent In this setting we consider a finite number of slip directions 
$\vec{b} \in \R^2$ and to  each $\vec{b}$ we will associate a
dislocation density. For a detailed  physical presentation of a model with
multi-slip directions, we  refer to Yefimov, Van der Giessen   
\cite{Yef} and  Yefimov \cite[ch. 5.]{Yef1}  and to  Groma, Balogh 
\cite{Groma} for the case of a model with a single slip direction . See
also Cannone et al. \cite{EC} for a mathematical analysis of the
Groma, Balogh model. In Subsection \ref{EM:mod_2D}, we present the 2D-model
with multi-slip directions. \\
  
\noindent  In the particular geometry where the dislocations densities
only depend  on the  variable $x=x_1+x_2$, this two-dimensional model reduces to
one-dimensional model which presented in In Subsection \ref{EM:mod_1D}. See El Hajj \cite{EL} and El Hajj, Forcadel
\cite{EF} for a  study in the special case of a single slip
direction. Finally in  Subsection \ref{EM:mod_1D_np}, we explain how to
recover equation (\ref{EM:burger}) as a model for dislocation dynamics with 
$\d{a^i(u)=\sum_{j=1,\dots,M}A_{ij}u^j}$ for some particular
non-negative and symmetric matrix $A$.

\subsection{The 2D-model}\label{mod_2D}

We now present in details the  two-dimensional model. We denote by ${\bf
  X}$ the vector ${\bf X} = (x_1,x_2)$. We consider a crystal filling
the whole space $\R^2$ and its displacement
$v=(v_1,v_2):\R^2\rightarrow\R^2$, where we have not yet introduced the time
dependence for the moment.\\ 

\noindent We define the total strain by
$$\varepsilon(v)=\frac{1}{2}(\nabla v+{}^t\nabla v),$$

\noindent where $\nabla v$ is the gradient with $\displaystyle{(\nabla v)_{ij}
  = \frac{\partial v_i}{\partial x_j}},\;\; i,j \in \{1,2\}$.\\

\noindent Now, we assume that the dislocations densities under consideration
are associated to edge dislocations. This means that  we
consider $M$ slip directions where each direction is caraterize by a
Burgers vectors $\vec{b}^k=(b_1^k,b_2^k)\in \R^2$, for $k=1,\dots,M$. This leads
to $M$ type of dislocations which  propagate in the plan $(x_1,x_2)$ following
the direction of $\vec{b}^k$, for $k=1,\dots,M$.\\

\noindent The total strain can be splitted in two parts:

$$
\varepsilon(v) = \e^e + \e^p.
$$

\noindent Here, $\e^e$ is the elastic strain and $\e^p$ the plastic 
strain defined by
\begin{equation}\label{EM:eq:epsilon0}
\varepsilon^p=\sum_{k=1,\dots, M}\varepsilon^{0,k} u^k,
\end{equation}

\noindent where, for each $k=1,\dots,M$,  the scalar function $u^k$  is the
plastic displacement associated to the $k$-th slip system whose 
matrix $\varepsilon^{0,k}$ is defined by

\begin{equation}\label{EM:eq:e0}\varepsilon^{0,k}=\frac{1}{2}\left(\vec{b}^k\otimes\vec{n}^{k}
  +\vec{n}^{k}\otimes \vec{b}^k\right),\end{equation} 

\noindent where $\vec{n}^{k}$ is unit  vector orthogonal  to $\vec{b}^k$ and 
$\left(\vec{b}^k\otimes\vec{n}^{k}\right)_{ij}=b_i^k n^{k}_j$.\\

\noindent To simplify the presentation, we assume the simplest possible
periodicity property of the unknowns.\\

{\it \noindent \underline{Assumption $(H)$}:

\noindent i) The function $v$ is $\Z^2$-periodic with $\d{\int_{(0,1)^2}v\;
  d{\bf X}=0}.$ \\

\noindent ii)  For each $k=1,\dots,M$, there exists $L^k\in\R
^2$ such that $u^k-L^k\cdot{\bf X}$ is a $\Z^2$-periodic.\\

\noindent iii) The integer $M$ is even with $M=2N$ and  $L^{k+N}=L^k$, and that

$$L^{k+N}=L^k,\;\;\vec{b}^{k+N}=-\vec{b}^{k},\;\; \vec{n}^{k+N}=\vec{n}^{k},$$
$$\varepsilon^{0,k+N}=-\varepsilon^{0,k}.$$

\noindent iv) We 
denote by $\vec{\tau}^k=(\tau^k_1,\tau^k_2)$ a vector parallel to
$\vec{b}^{k}$ such that $\vec{\tau}^{k+N}=\vec{\tau}^{k}$. We require that $L^k$ is
chosen such  $\vec{\tau}^{k}\cdot L^k \ge 0$.}\\

\noindent The plastic displacement $u^k$ is related to the dislocation
density associated to the Burgers vector $\vec{b}^{k}$. We have 

\begin{equation}\label{EM:posi0}
\mbox{$k$-th dislocation density$\;\;=\vec{\tau}^{k}\cdot \nabla
  u^k\ge0$}.\end{equation}

\noindent The stress is then given by 
\begin{equation}\label{EM:eq:sigma}
\sigma=\Lambda :\varepsilon^e,
\end{equation}

\noindent {\it i.e.} the coefficients of the matrix $\sigma$ are:
$$
\sigma_{ij} = \sum_{k,l=1,2} \Lambda_{ijkl}
\varepsilon^e_{kl}\quad\mbox{for}\quad i,j=1,2,
$$

\noindent where $\d{\Lambda=\left(\Lambda_{ijkl}\right)_{i,j,k,l=1,2}}$,
are the constant  elastic 
 coefficients of the material, satisfying for $m>0$:
\begin{equation}\label{EM:coercivite}
\sum_{ijkl=1,2} \Lambda_{i,j,k,l}\varepsilon_{ij}\varepsilon_{kl}\geq m\sum_{i,j=1,2}\varepsilon_{ij}^2
\end{equation}
for all symmetric matrices $\varepsilon=\left(\varepsilon_{ij}\right)_{ij}$,
{\it i.e.} such that $\varepsilon_{ij}=\varepsilon_{ji}.$\\

\noindent Finally, for $k=1,\dots,M$,  the functions $u^k$ and $v$ are
then assumed to depend on $(t,{\bf X})\in (0,T)\times \R^2$ and to be solutions of the coupled
system (see  Yefimov \cite[ch. 5.]{Yef1} and Yefimov, Van der Giessen \cite{Yef}):

\begin{center}\begin{equation}\label{EM:eq:elasdis}\left\{\begin{array} {lll}
 \displaystyle{\div\;\sigma}& = 0&\mbox{ on $
   (0,T)\times \R^2$},\\
\sigma&=\Lambda :\left(\varepsilon(v)-\varepsilon^p\right)& \mbox{ on $
   (0,T)\times \R^2$},\\
\varepsilon(v) &=\frac{1}{2}\left(\nabla v+{}^t\nabla
                  v\right)& \mbox{ on $
   (0,T)\times \R^2$},\\
\varepsilon^p&=\displaystyle{\sum_{k=1,\dots,
    M}\varepsilon^{0,k}u^k}
& \mbox{ on $
   (0,T)\times \R^2$},\\
$\;$\\
\displaystyle{\partial_t u^k} &
=(\sigma:\varepsilon^{0,k})\displaystyle{\vec{\tau}^k.\nabla u^k} 
& \mbox{ on $
   (0,T)\times \R^2$,\quad for $k=1,\dots,M$}, \\
\end{array} \right.\end{equation}\end{center}

\noindent {\it i.e.} in  coordinates
\begin{center}\begin{equation}\label{EM:coord_eq:elasdis}\left\{\begin{array} {lll}
\displaystyle{\sum_{j=1,2}\frac{\partial \sigma_{ij}}{\partial x_j}} =
0 \hspace{3.5cm}\mbox{ on $(0,T)\times \R^2$, \hspace{0.3cm}\quad for $i=1,2$},\\

\left.\begin{array} {lll}
\sigma_{ij} = \displaystyle{\sum_{k,l=1,2}} \Lambda_{ijkl}\left
  (\varepsilon_{kl}(v)-
\varepsilon^p_{kl}\right)& \mbox{ on $
   (0,T)\times \R^2$},\\

\varepsilon_{ij}(v) =\d{\frac{1}{2}\left(\frac{\partial v_i}{\partial
    x_j}+
\frac{\partial v_j}{\partial x_i}\right)}& \mbox{ on $
   (0,T)\times \R^2$},\\

\varepsilon_{ij}^p=\displaystyle{\sum_{k=1,\dots,
    M}\varepsilon^{0,k}_{ij}u^k}
& \mbox{ on $
   (0,T)\times \R^2$},
\end{array}\right|\mbox{\quad for $i,j=1,2$} \\
\\
\displaystyle{\partial_t u^k} 
=\left( \displaystyle{\sum_{i,j\in\{1,2\}}
    \sigma_{ij}}\varepsilon_{ij}^{0,k}\right)\displaystyle{\vec{\tau}^k.\nabla
u^k}  \hspace{0.3cm}\mbox{ on $
   (0,T)\times \R^2$, \hspace{0.3cm}\quad for $k=1,\dots,M$},
\end{array} \right.
\end{equation}\end{center}

\noindent where the unknowns of the system are $u^k$ and the displacement $v
= (v_1,v_2)$ and with $\varepsilon^{0,k}$  defined in
(\ref{EM:eq:e0}). Here the first equation of  (\ref{EM:eq:elasdis}) is the
equation of elasticity, while the last equation of (\ref{EM:eq:elasdis}) is
the transport equation satisfied by the plastic displacement whose
velocity is given by the Peach-Koehler force
$\sigma:\varepsilon^{0,k}$. Remark that this implies in particular that
each dislocation density satisfies a conservation law (see the equation
obtained by derivation, using (\ref{EM:posi0})). Remark also that our
equations are compatible with our periodicity assumptions $(H)$, $(i)$-$(ii)$. \\
\subsection{Derivation of the 1D-model}\label{EM:mod_1D}

\noindent In this Subsection we are interested in a particular geometry where the dislocations  densities
depend only on the  variable $x=x_1+x_2$. This will lead to
1D-model. More precisely, we make the following:\\

{\it \noindent \underline{Assumption $(H')$}:

\noindent i) The functions $v(t, {\bf X})$ and  $u^k(t, {\bf
  X})-L^k\cdot{\bf X}$  depend on the variable $x=x_1+x_2$.\\

\noindent ii)  $\tau^k_1+\tau^k_2=1,$ for $k=1,\dots,M$.\\

\noindent iii) $L_1^k=L_2^k$ for $k=1,\dots,M$.}\\

\noindent For this particular one-dimensional  geometry, we denote by an
abuse of notation the function $v=v(t,x)$ which is $1$-periodic in
$x$. If we set $l^k=\frac {L_1^k+L_2^k}{2}$, we have 
$$L^k\cdot{\bf X}=l^k\cdot x+\left(\frac{L_1^k-L_2^k}{2}\right)(x_1-x_2).$$

\noindent By assumption $(H')$, $(iii)$, we see (again  by an abuse of
notation) that $\d{u=(u^k(t,x))_{k=1,\dots,M}}$ is such that
for $k=1,\dots,M$, $u^k(t,x)-l^k\cdot x$  is  $1$-periodic in $x$.\\

\noindent  Now, we can integrate the equations of elasticity, {\it i.e.}
the first equation of (\ref{EM:eq:elasdis}). Using the
$\Z^2$-periodicity of the  unknowns (see assumption $(H)$, $(i)$-$(ii)$),
and the fact that $\varepsilon^{0,k+N}=-\varepsilon^{0,k}$ (see
assumption $(H)$, $(iii)$), we can easily conclude that  the strain 

\begin{equation}\label{EM:strain}\mbox{$\e^e$ as a linear function of
$\d{(u^j-u^{j+N})_{j=1,\dots,N}}$ \quad and of
$\d{\left(\int_0^1(u^j-u^{j+N})\ dx\right)_{j=1,\dots,N}}$.}
\end{equation}

\noindent This leads to the following Lemma 

\begin{lem}{\bf (Stress for the 1D-model)}\label{EM:1D}\\
Under assumptions $(H)$, $(i)$-$(ii)$-$(iii)$ and $(H')$, $(i)$-$(iii)$
and (\ref{EM:coercivite}), we have 

\begin{equation}\label{EM:stress}-\sigma:\varepsilon^{0,i}=\sum_{j=1,\dots,M}A_{ij}u^j+
\sum_{j=1,\dots,M}Q_{ij}\int_0^1u^j\ dx, \quad \mbox{for $i=1,\dots,N$}.
\end{equation}

\noindent where for $i,j=1,\dots,N$

\begin{equation}\label{EM:matrix}\left\{\begin{array}{ll}
&\mbox{$A_{i,j}=A_{j,i}$ \quad and\quad
  $A_{i+N,j}=-A_{i,j}=A_{i,j+N}$,}\\
\\
&\mbox{$Q_{i,j}=Q_{j,i}$ \quad and \quad
  $Q_{i+N,j}=-Q_{i,j}=Q_{i,j+N}$.}
\end{array}\right.\end{equation}

\noindent Moreover the matrix $A$ is non-negative.
\end{lem}

\noindent The proof of Lemma \ref{EM:1D} will be given at the end of this
Subsection.\\

\noindent Finally using Lemma \ref{EM:1D}, we can eliminate the stress and  reduce
the problem  to a one-dimensional system of $M$ transport equations only
depending on the function  $u^i$, for
$i=1,\dots,M$. Naturally, from (\ref{EM:stress}) and $(H')$, $(ii)$ this 1D-model
has the following form

\begin{equation}\label{EM:burger_loc}
\partial_t
u^i+\left(\sum_{j=1,\dots,M}A_{ij}u^j+\sum_{j=1,\dots,M}Q_{ij}\int_0^1u^j\ dx\right)
\partial_x u^i=0,\qquad \mbox{on $(0,T)\times \R$, $\;\;$for $i=1,\dots,M$},
\end{equation}

\noindent with from (\ref{EM:posi0})

\begin{equation}\label{EM:croissante}\partial_x u^i\ge 0 \quad \mbox{for
    $i=1,\dots,M$.}
\end{equation}

\noindent Now, we give the proof of Lemma \ref{EM:1D}.\\

\noindent {\bf Proof of Lemma \ref{EM:1D}:}\\
\noindent For the 2D-model, let us consider the elastic energy on the periodic cell
(using the fact that $\e^e$ is $\Z^2$-periodic)

$$E^{el}=\frac 12\int_{(0,1)^2} \Lambda:\e^e:\e^e\ d{\bf X}.$$

\noindent By definition of $\sigma$ and $\e^e$, we have for
$i=1,\dots,M$

\begin{equation}\label{EM:energy}
\sigma:\varepsilon^{0,i}=-\nabla_{u^i}E^{el}.
\end{equation}

\noindent On the other hand usind $(H')$, $(i)$-$(iii)$, (with $x=x_1+x_2$) we can check that we can
rewrite  the elastic energy as 
$$E^{el}=\frac 12\int_{0}^1 \Lambda:\e^e:\e^e\ dx.$$

\noindent Replacing $\e^e$ by its expression (\ref{EM:strain}), we get:

 $$\begin{array}{ll}\d{E^{el}}
=
&\d{\frac
  12\int_0^1\sum_{i,j=1,\dots,N}A_{ij}(u^j-u^{j+N})(u^i-u^{i+N}) \ dx }\\
\\
&+
\d{\frac12\sum_{i,j=1,\dots,N}Q_{ij}\left(\int_0^1(u^j-u^{j+N})\ dx\right)
\left(\int_0^1(u^i-u^{i+N})\ dx\right),}\end{array}$$

\noindent for some symmetric matrices $A_{i,j}=A_{j,i}$,
$Q_{i,j}=Q_{j,i}$. In particular, joint to (\ref{EM:energy}) this gives
exactly (\ref{EM:stress}) with (\ref{EM:matrix}).\\

\noindent Let us now consider the  functions $w^i=u^i-u^{i+N}$ such
that 

\begin{equation}\label{EM:integ}\d{\int_0^1w^i \ dx}=0 \quad \mbox{for
    i=1,\dots,N,}\end{equation}

\noindent From (\ref{EM:coercivite}) that we deduce that

$$\d{0\le E^{el}=\frac
12\int_0^1\sum_{i,j=1,\dots,N}A_{ij}w^iw^j \ dx}.$$

\noindent More precisely, for all
$i=1,\dots,N$ and for all $\bar{w}^i\in \R$, we set 

$$w^i=\left\{\begin{array}{ll}
\bar{w}^i& \quad \mbox{on}\quad [0,\frac 12],\\
-\bar{w}^i & \quad \mbox{on}\quad [\frac 12,1],\\
\end{array}\right.$$

\noindent which satisfies (\ref{EM:integ}). Finally, we obtain that  
$$0\le E^{el}=\frac
12\int_0^1\sum_{i,j=1,\dots,N}A_{ij}\bar{w}^i\bar{w}^j\ dx.$$

\noindent Because this is true for every $\bar{w}^i$, we deduce that $A$
a non-negative matrix. $\hfill\Box$

\subsection{Heuristic derivation of the non-periodic model}\label{EM:mod_1D_np}

\noindent Starting from the model (\ref{EM:burger_loc})-(\ref{EM:croissante})
where for $i=1,\dots,M,$, $u^i(t,x)-l^i\cdot x$  is  $1$-periodic in $x$, we now want to
rescale the  unknowns to make the periodicity disappear. More precisely,
we have the following Lemma:

\begin{lem}{\bf (Non-periodic model)}\label{EM:non_per}\\
Let $u$ be a solution of  (\ref{EM:burger_loc})-(\ref{EM:croissante}) assuming Lemma
\ref{EM:1D} and $u^i(t,x)-l^i\cdot x$  is  $1$-periodic in $x$. Let

$$u_{\delta}^j(t,x)=u^j(\delta t,\delta x),\quad \mbox{for a small 
$\delta>0$ and  for $j=1,\dots,M$,}
$$

\noindent such that, for all $j=1,\dots,M$

\begin{equation}\label{EM:limit}u_{\delta}^j(0,\cdot)\to \bar{u}^j(0,\cdot),
  \quad\mbox{as}\quad \delta\to 0, \quad\mbox{and}\quad
\bar{u}^j(0,\pm \infty)=\bar{u}^{j+N}(0,\pm \infty)\end{equation}

\noindent Then $\d{\bar{u}=(\bar{u}^j)_{j=1,\dots,M}}$ formally is a
solution of

\begin{equation}\label{EM:limit_p}\partial_t
\bar{u}^{i}+\left(\sum_{j=1,\dots,M}A_{ij}\bar{u}^{j}\right)
\partial_x \bar{u}^{i}=0,\qquad \mbox{on $(0,T)\times \R$},\end{equation}

\noindent with  the matrix $A$ is non-negative and $\partial_x
\bar{u}^{i} \ge 0$ for $i=1,\dots,M$.

\end{lem}

\noindent  We remark that the limit problem (\ref{EM:limit_p}) is of type
(\ref{EM:burger}) with $(H1')$ and $(H2')$.\\
 
\noindent Now, we give a formal proof of  Lemma \ref{EM:non_per}.\\

\noindent {\bf  Formal proof of Lemma \ref{EM:non_per}:}\\
\noindent Here, we know that  $u_{\delta}^i-\delta l^i\cdot x$ is  $\d{\frac
{1}{\delta}}$-periodic in $x$, and satisfies for $i=1,\dots,M$

\begin{equation}\label{EM:limit_p1}\partial_t
u_{\delta}^i+\left(\sum_{j=1,\dots,M}A_{ij}u_{\delta}^j+\delta\sum_{j=1,
\dots,M}Q_{ij}\int_0^{\frac
      {1}{\delta}}u_{\delta}^j\ dx\right)
\partial_x u_{\delta}^i=0,\qquad \mbox{on $(0,T)\times \R$,}
\end{equation}

\noindent To simplify, assume that the initial data
$u_{\delta}(0,\cdot)$ converge to a function $\bar{u}(0,\cdot)$ such
that $\partial_x u_{\delta}(0,\cdot)$ has a support in $(-R,R)$, uniformly
in $\delta$, where $R$ a positve constant. We expect heuristically that the velocity in
(\ref{EM:limit_p1}) remains uniformly bounded  as $\delta\to 0$.\\

\noindent Therefore, using the finite propagation speed, we see that,  
 there exists a constant $C$ independent in $\delta$, such that 
$\partial_x u_{\delta}(t,\cdot)$ has a support in $(-R-Ct,R+Ct)$
uniformly in  $\delta$. Moreover, from (\ref{EM:limit}) and the fact that 

 $$\sum_{j=1,\dots,M}Q_{ij}\int_0^{\frac
      {1}{\delta}}u_{\delta}^j\ dx=\sum_{j=1,\dots,N}Q_{ij}\int_0^{\frac
      {1}{\delta}}(u^j-u^{j+N})\ dx,$$
\noindent we deduce that 

$$\sum_{j=1,\dots,M}Q_{ij}\int_0^{\frac
      {1}{\delta}}u_{\delta}^j\ dx,$$

\noindent  remains  bounded uniformly in  $\delta$. Then formally the 
non-local term vanishes and we get for  $i=1,\dots,M$

$$\sum_{j=1,\dots,M}A_{ij}u_{\delta}^j+ \delta\sum_{j=1,
\dots,M}Q_{ij}\int_0^{\frac {1}{\delta}}u_{\delta}^j\ dx \to
\sum_{j=1,\dots,M}A_{ij}\bar{u}^j,\quad\mbox{as}\quad \delta\to 0,$$

\noindent which proves that $\bar{u}$ is solution of (\ref{EM:limit_p}),
with the matrix $A$ is non-negative . $\hfill\Box$

\section{Appendix: proof of Theorem \ref{EM:unicite}}

Let $u_1=(u_1^i)_i$ and  $u_2=(u_2^i)_i$, for $i=1,\cdots,M$, be two solutions of
the system (\ref{EM:burger}) in ${[W^{1,\infty}((0,T)\times
  \mathbb{R})]}^M$, such that $u_1^i(0,\cdot)=u_2^i(0,\cdot)$.\\

\noindent Then by definition $u_1^i$ and  $u_2^i$ satisfy
respectively  the following system, for $i=1,\cdots,M$:

$$\partial_t u^i_1=-a^i(u_1)\partial_x u^i_1,$$

$$\partial_t u^i_2=-a^i(u_2)\partial_x u^i_2,$$
\noindent Subtracting the two equations we get:
$$\partial_t \left(u^i_1-u^i_2\right)=
-\left(a^i(u_1)-a^i(u_2)\right)\partial_x u^i_1
-a^i(u_2)\partial_x(u^i_1-u^i_2).
$$
 
\noindent Multiplying this system by
$\left(u^i_1-u^i_2\right)(\psi)^2$ where $\psi(x)=e^{-|x|}$,
and integrating in space, we deduce
that:
$$\begin{array}{ll}\displaystyle{\frac 12\frac
    {d}{dt}\left\|(u^i_1-u^i_2)
\psi\right\|_{L^2(\R)}^2}=
&\displaystyle{-\int_{\R}\left(a^i(u_1)-a^i(u_2)\right)\left(u^i_1-u^i_2\right)\psi^2\partial_x u^i_1}\\
\\
&\displaystyle{-\int_{\R}a^i(u_2)\psi^2\left(u^i_1-u^i_2\right)\partial_x(u^i_1-u^i_2).}
\end{array}$$
\noindent  Taking the sum over $i$, we get: 
$$\begin{array}{ll}\displaystyle{\frac 12 \frac
{d}{dt}\left(\sum_{i=1,\dots ,M}\left\|(u^i_1-u^i_2)\psi
\right\|_{L^2(\R)}^2\right)}=
&\overbrace{ \mathstrut\displaystyle{-\int_{\R}\sum_{i=1,\dots ,M}
\left(a^i(u_1)-a^i(u_2)\right)\left(u^i_1-u^i_2\right)\psi^2\partial_x u^i_1}}^{I_1}\\
\\
&\overbrace{ \mathstrut\displaystyle{-\frac 12\int_{\R}\sum_{i=1,\dots
      ,M}a^i(u_2)\psi^2\partial_x(u^i_1-u^i_2)^2}}^{I_2}.
\end{array}$$
\noindent Integrating  $I_2$ by part,  we obtain:

$$\begin{array}{ll}I_2=
&\overbrace{ \mathstrut\displaystyle{\frac 12\int_{\R}\sum_{i,j=1,\dots ,M}
a^i_{,j}(u_2)(\partial_xu^j_2)\psi^2(u^i_1-u^i_2)^2}}^{I_{21}}\\
\\
&+\overbrace{ \mathstrut\displaystyle{\frac 12\int_{\R}\sum_{i=1,\dots
      ,M}a^i(u_2) (u^i_1-u^i_2)^2 
\partial_x(\psi^2)}}^{I_{22}}.
\end{array}$$
\noindent Next, using the fact that $u_2^i$ is bounded in $W^{1,\infty}((0,T)\times
\mathbb{R})$, for $i=1,\dots,M$, we deduce that:

\begin{equation}\begin{array}{ll}\label{EM:I21}\left|I_{21}\right|
&\le \frac 12 MM_1\|u_2\|_{[W^{\infty}((0,T)\times\R)]^M}
\displaystyle{\left(\sum_{i=1,\dots,M}\left\|(u^i_1-u^i_2)\psi\right\|_{L^2(\R)}^2\right),}\\
\\
&\le C
\displaystyle{\left(\sum_{i=1,\dots,M}\left\|(u^i_1-u^i_2)\psi\right\|_{L^2(\R)}^2\right).}
\end{array}\end{equation}

\noindent Since $\partial_x(\psi(x))^2=-2sign(x)(\psi(x))^2$ and $u_2^i$
is bounded in $W^{1,\infty}((0,T)\times \mathbb{R})$, for
$i=1,\cdots,M$, we obtain:

\begin{equation}\begin{array}{ll}\label{EM:I22}\left|I_{22}\right|
&\le \frac 12 M_0
\displaystyle{\left(\sum_{i=1,\dots,M}\left\|(u^i_1-u^i_2)\psi\right\|_{L^2(\R)}^2\right)}\\
\\
&\le C
\displaystyle{\left(\sum_{i=1,\dots,M}\left\|(u^i_1-u^i_2)\psi\right\|_{L^2(\R)}^2\right)}
\end{array}\end{equation}

\noindent Now, using the fact that  $u_1^i$ is bounded in $W^{1,\infty}((0,T)\times
\mathbb{R})$, for $i=1,\cdot,\cdot,M$, and
the inequality $|ab|\le \frac 12 (a^2+b^2)$, we get:
\begin{equation}\begin{array}{ll}\label{EM:I1}
\left|I_1\right|
&\displaystyle{\le \frac{1}{2}M_1 (M+1)\|u_1\|_{[W^{\infty}((0,T)\times\R)]^M}
\displaystyle{\int_{\R}\sum_{i=1,\dots,M}|u^i_1-u^i_2|^2\psi^2},}\\
\\
&\displaystyle{\le\frac{1}{2} M_1 (M+1)\|u_1\|_{[W^{\infty}((0,T)\times\R)]^M}
\left(\sum_{i=1,\dots,M}\left\|(u^i_1-u^i_2)\psi\right\|_{L^2(\R)}^2\right),}\\
\\
&\le C
\displaystyle{\left(\sum_{i=1,\dots,M}\left\|(u^i_1-u^i_2)\psi\right\|_{L^2(\R)}^2\right).}
\end{array}\end{equation}
\noindent  Finally,  (\ref{EM:I1}), (\ref{EM:I21})  and (\ref{EM:I22}), imply:
$$\displaystyle{\frac
{d}{dt}\left(\sum_{i=1,\dots,M}\left\|(u^i_1-u^i_2)\psi
\right\|_{L^2(\R)}^2\right)}
\le 2\left(\left|I_1\right|+ \left|I_{21}\right|+\left|I_{22}\right|\right)\le  C
\left(\sum_{i=1,\dots,M}\left\|(u^i_1-u^i_2)\psi\right\|_{L^2(\R)}^2
\right).$$
\noindent Now, we apply the Gronwall Lemma 
and we use that
$u_1^i(0,\cdot)=u_2^i(0,\cdot)$, to deduce that:

$$\displaystyle{\sum_{i=1,\dots,M}\left\|(u^i_1-u^i_2)\psi\right\|_{L^{\infty}((0,T);
  L^2(\R))}^2\le \sum_{i=1,\dots,M}\left\|\left(u^i_1(0,\cdot)-u^i_2(0,\cdot)\right)\psi\right\|_{
  L^2(\R)}^2 e^{CT}=0,}$$
\noindent i.e., $u_1=u_2$ a.e in $(0,T)\times \R$. 
$\hfill\Box$
\section{Acknowledgements }
The first author would like to thank M. Cannone and M. Jazar
 for fruitful remarks that helped in the preparation of the paper. This work was partially
supported by the contract JC 1025 ``ACI,
jeunes chercheuses et jeunes chercheurs'' (2003-2007) and the program ``PPF, programme pluri-formations mathématiques
financières et EDP'', (2006-2010), Marne-la-Vall\'ee University and École Nationale
des Ponts et Chaussées.

\bibliographystyle{siam}
\bibliography{biblio}
\end{document}